\documentclass[letterpaper,twocolumn,10pt,anonymous]{article}
\usepackage{usenix2019_v3}

\usepackage{tikz}
\usetikzlibrary{arrows, calc, positioning, decorations.markings, decorations.pathreplacing} 
\usepackage{amsmath}

\usepackage[binary-units]{siunitx}
\DeclareSIUnit{\billion}{\text{billion}}
\usepackage{nicefrac}

\usepackage{pgfplots}                                                           
\usepackage{pgf}                                                           
\usepgfplotslibrary{statistics}
\usepackage{csvsimple}

\usepackage{balance}

\usepackage{adjustbox}
\pgfplotsset{compat=1.13}

\usepackage{xspace}
\newcommand{\fuzztool}{\emph{Frankenstein}\xspace}

\usepackage{subfig}
\usepackage{xcolor}
\usepackage{xifthen}
\newcommand{\ifequals}[3]{\ifthenelse{\equal{#1}{#2}}{#3}{}}

\definecolor{darkred}{rgb}{0.831, 0, 0.063}
\definecolor{darkgreen}{rgb}{0.043, 0.768, 0.003}
\definecolor{darkblue}{rgb}{0, 0.2, 0.6}

\newcommand{\cvea}{\emph{CVE-2019-11516}\xspace}
\newcommand{\cveb}{\emph{CVE-2019-13916}\xspace}
\newcommand{\cvec}{\emph{CVE-2019-15063}\xspace}
\newcommand{\cved}{\emph{CVE-2019-18614}\xspace}

\usepackage[firstpage]{draftwatermark}
\SetWatermarkText{\hspace*{6in}\raisebox{7.5in}{\includegraphics{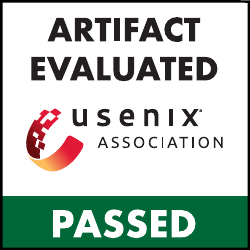}}}
\SetWatermarkAngle{0}

\usepackage{listings} 
\lstdefinelanguage{ASM}{
    morekeywords={b, ble, blt, bne, bx, bl, ldr, str, push, pop, mov, add, sub},
    keywordstyle=\color{blue},
    sensitive=false, 
    morecomment=[l]{//}, 
    morecomment=[s]{/*}{*/}, 
    morestring=[b]", 
} %
\lstdefinelanguage{none}{
  identifierstyle=
}

\usepackage[nolist]{acronym}
\begin{acronym}
\acro{SDR}{Software-Defined Radio}
\acro{AGC}{Automatic Gain Control}
\acro{GCI}{Global Coexistence Interface}
\acro{ECI}{Enhanced Coexistence Interface}
\acro{AFH}{Adaptive Frequency Hopping}
\acro{FEM}{Front-End Module}
\acro{SP3T}{Single Pole, Triple Throw}
\acro{HCI}{Host Controller Interface}
\acro{A2DP}{Advanced Audio Distribution Profile}
\acro{SCO}{Synchronous Connection-Oriented}
\acro{GPIO}{General Purpose Input Output}
\acro{ASLR}{Address Space Layout Randomization}
\acro{LMP}{Link Management Protocol}
\acro{LCP}{Link Control Protocol}
\acro{LNA}{Low-Noise Amplifier}
\acro{EIR}{Extended Inquiry Response}
\acro{CRC}{Cyclic Redundancy Check}
\acro{RTOS}{Real-Time Operating System}
\acro{QEMU}{Quick Emulator}
\acro{UART}{Universal Asynchronous Receiver Transmitter}
\acro{MITM}{Machine-in-the-Middle}
\acro{MAC}{Media Access Control}
\acro{BLE}{Bluetooth Low Energy}
\acro{ACL}{Asynchronous Connection-Less}
\acro{DSP}{Digital Signal Processing}
\acro{SDIO}{Secure Digital Input Output}

\acro{BCS}{Bluetooth Core Scheduler}
\acro{ELF}{Executable and Linking Format}
\acro{GIAC}{Global Inquiry Access Code}
\acro{JSON}{JavaScript Object Notation}
\acro{L2CAP}{Logical Link Control and Adaptation Protocol}
\acro{LMP}{Link Management Protocol}
\acro{PTM}{Pseudo Terminal Master}
\acro{PTS}{Pseudo Terminal Slave}
\acro{QEMU}{Quick Emulator}
\acro{RCE}{Remote Code Execution}
\acro{RFU}{Reserved for Future Use}
\acro{SVC}{Supervisor Call}
\acro{UART}{Universal Asynchronous Receiver Transmitter}
\acro{WICED}{Wireless Internet Connectivity for Embedded Devices}
\acro{MMIO}{Memory Mapped Input/Output}
\acro{LE}{Bluetooth Low Energy}
\acro{IoT}{Internet of Things}
\acro{IDE}{Integrated Development Environment}
\acro{ARM}{Advanced RISC Machine}
\acro{LM}{Link Manager}

\acro{RTOS}{Real-Time Operating System}
\acro{DMA}{Direct Memory Access}
\acro{RXDMA}{Receive Direct Memory Access}
\acro{NVRAM}{Non-Volatile Random-Access Memory}
\acro{ROM}{Read-Only Memory}
\acro{Rx}{Receive}
\acro{Tx}{Transmit}
\acro{SPI}{Serial Peripheral Interface}
\acro{MSP}{Main Stack Pointer}
\acro{PSP}{Process Stack Pointer}
\acro{RF}{Radio Frequency}
\acro{BLOB}{Binary Large Object}
\acro{SCO}{Synchronous Connection-Oriented}
\acro{UAF}{Use-After-Free}
\acro{FHS}{Frequency Hop Sync}
\acro{CRC}{Cyclic Redundancy Check}
\acro{PDU}{Protocol Data Unit}
\acro{NFC}{Near Field Communication}
\acro{JPEG}{Joint Photographic Experts Group}
\acro{LPE}{Local Privilege Escalation}
\acro{DEP}{Data Execution Prevention}
\acro{XN}{eXecute Never}
\acro{LCP}{Link Control Protocol}
\acro{GATT}{Generic Attribute}
\acro{PoC}{Proof of Concept}
\newacroplural{PoC}[PoCs]{Proofs of Concept}
\acro{EDR}{Enhanced Data Rate}
\acro{MWS}{Mobile Wireless Standards}
\acro{HAL}{Hardware Abstraction Layer}
\acro{LR}{Link Register}
\acro{eSIM}{embedded-SIM}
\acro{RSP}{Remote SIM Provisioning}
\acro{PC}{Program Counter}
\acro{MD}{More Data}
\acro{LLID}{Logical Link Identifier}
\acro{SFI}{Serial Flash Interface}
\acro{EEPROM}{Electrically Erasable Programmable Read-Only Memory}
\acro{TOFU}{Trust On First Use}
\acro{CVE}{Common Vulnerabilities and Exposure}
\acro{PCIe}{Peripheral Component Interconnect Express}
\acro{CFI}{Control Flow Integrity}
\acro{DoS}{Denial of Service}
\acro{MIPS}{Microprocessor without Interlocked Pipelined Stages}
\acro{CAN}{Controller Area Network}

\end{acronym}

\usepackage[colorinlistoftodos,prependcaption,disable]{todonotes} 
\presetkeys%
    {todonotes}%
    {inline,backgroundcolor=orange}{}

\begin{document}

\date{}


\title{\Large \bf Frankenstein: Advanced Wireless Fuzzing to \\Exploit New Bluetooth Escalation Targets}



\author{
{\rm Jan Ruge}\\
Secure Mobile Networking Lab\\
TU Darmstadt
\and
{\rm Jiska Classen}\\
Secure Mobile Networking Lab\\
TU Darmstadt
\and
{\rm Francesco Gringoli}\\
Dept. of Information Engineering\\
University of Brescia
\and
{\rm Matthias Hollick}\\
Secure Mobile Networking Lab\\
TU Darmstadt
} 

\pagestyle{empty}

\maketitle

\begin{abstract}
Wireless communication standards and implementations have a troubled history regarding security.
Since most implementations and firmwares are closed-source, fuzzing remains one of the main methods to uncover \ac{RCE} vulnerabilities in deployed systems.
Generic over-the-air fuzzing suffers from several shortcomings, such as constrained speed, limited repeatability, and restricted ability to debug.
In this paper, we present \fuzztool, a fuzzing framework based on advanced firmware emulation, which addresses these shortcomings.
\fuzztool brings firmware dumps ``back to life'', and provides fuzzed input to the chip's virtual modem.
The speed-up of our new fuzzing method is sufficient to maintain interoperability with the attached operating system, hence triggering realistic full-stack behavior.
We demonstrate the potential of \fuzztool by finding three zero-click vulnerabilities in the \emph{Broadcom} and \emph{Cypress} Bluetooth stack, which is used in most \emph{Apple} devices, many \emph{Samsung} smartphones, the \emph{Raspberry Pis}, and many others.

Given \ac{RCE} on a Bluetooth chip, attackers may escalate their privileges beyond the chip's boundary.
We uncover a \mbox{Wi-Fi}/Bluetooth coexistence issue that crashes multiple operating system kernels and a design flaw in the Bluetooth 5.2 specification that allows link key extraction from the host.
Turning off Bluetooth will not fully disable the chip, making it hard to defend against \ac{RCE} attacks.
Moreover, when testing our chip-based vulnerabilities on those devices, we find \emph{BlueFrag}, a chip-independent \emph{Android} \ac{RCE}.

\end{abstract}


\tikzset{>=latex}

\newcommand{\blockSetColor}[1]{
    \def \blockColor {blue!30}
    \ifequals{#1}{Free}{\def \blockColor {green!30}}{}
    \ifequals{#1}{Free Head}{\def \blockColor {green!30}}{}
    \ifequals{#1}{Free (Head)}{\def \blockColor {green!30}}{}
    \ifequals{#1}{Corrupted}{\def \blockColor {red!30}}{}
}
\newcommand{\blockBasicSetup}[4]{
    \blockSetColor{#1}
    \filldraw[fill=\blockColor] (0, 0) rectangle node (hdr1) {} ++(0.5,0.5); 
    \filldraw[fill=\blockColor] (0.5,0) rectangle node (buff1) {#1} ++(2.5,0.5); 

    \blockSetColor{#2}
    \filldraw[fill=\blockColor] (3,0) rectangle node (hdr2) {} ++(0.5,0.5); 
    \filldraw[fill=\blockColor] (3.5,0) rectangle node (buff2) {#2} ++(2.5,0.5); 

    \blockSetColor{#3}
    \filldraw[fill=\blockColor] (6,0) rectangle node (hdr3) {} ++(0.5,0.5); 
    \filldraw[fill=\blockColor] (6.5,0) rectangle node (buff3) {#3} ++(2.5,0.5); 

    \blockSetColor{#4}
    \filldraw[fill=\blockColor] (9,0) rectangle node (hdr4) {} ++(0.5,0.5); 
    \filldraw[fill=\blockColor] (9.5,0) rectangle node (buff4) {#4} ++(2.5,0.5); 

    \draw[->] ($(hdr4) + ( 0.1,-0.25)$) -| ($(hdr4) + ( 0.1,-0.6 )$) -> ($(hdr4) + (2.75,-0.6 )$) ;
    \draw (0,-1.5) rectangle node (block) {\texttt{BLOC} Struct} ++(2.5,0.5); 
    
}
\newcommand{\blockFreeEdge}[2]{
    \draw[->] 
        ($(#1) + ( 0.1,-0.25)$) -| 
        ($(#1) + ( 0.1,-0.6 )$) |- 
        ($(#2) + (-0.1,-0.6 )$) -> 
        ($(#2) + (-0.1,-0.25)$);
}
\newcommand{\blockArrowUp}[2]{
    \draw[->] 
        ($(#1) + ( 0.1, 0.25)$) -|
        ($(#1) + ( 0.1, 0.6 )$) |-
        ($(#2) + (-0.1,-0.8 )$) ->
        ($(#2) + (-0.1,-0.25)$) 
        ;
}
\newcommand{\blockArrowUpInv}[2]{
    \draw[->] 
        ($(#1) + (-0.1, 0.25)$) -|
        ($(#1) + (-0.1, 0.6 )$) |-
        ($(#2) + (-0.1,-0.8 )$) ->
        ($(#2) + (-0.1,-0.25)$) 
        ;
}
\newcommand{\blockArrowDown}[2]{
    \draw[->] 
        ($(#1) + ( 0.1,-0.25)$) -|
        ($(#1) + ( 0.1,-0.8 )$) |-
        ($(#2) + (-0.1, 0.6 )$) ->
        ($(#2) + (-0.1, 0.25)$) ;
}
\newcommand{\blockArrowDownInv}[2]{
    \draw[->] 
        ($(#1) + ( 0.1,-0.25)$) -|
        ($(#1) + ( 0.1,-0.8 )$) |-
        ($(#2) + ( 0.1, 0.6 )$) ->
        ($(#2) + ( 0.1, 0.25)$) ;
}

\newcommand{\BlockOverflowImageInitEmpty}{
\begin{tikzpicture}
    \blockBasicSetup{Free (Head)}{Free}{Free}{Free}

    \blockFreeEdge{hdr1}{hdr2}
    \blockFreeEdge{hdr2}{hdr3}
    \blockFreeEdge{hdr3}{hdr4}
    \blockArrowUp{block}{hdr1}

\end{tikzpicture}
}

\newcommand{\BlockOverflowImageInitAlloc}{
\begin{tikzpicture}
    \blockBasicSetup{Allocated}{Free (Head)}{Free}{Free}

    \blockFreeEdge{hdr2}{hdr3}
    \blockFreeEdge{hdr3}{hdr4}
    \blockArrowUp{block}{hdr2}
    \blockArrowDown{hdr1}{block}

\end{tikzpicture}
}

\newcommand{\BlockOverflowImageOvf}{
\begin{tikzpicture}
    \blockBasicSetup{Affected}{Corrupted}{Free}{Free}
    \draw (5,-1.5) rectangle node (target) {Target} ++(2.5,0.5); 

    \blockArrowDown{hdr2}{target}
    \blockFreeEdge{hdr3}{hdr4}
    \blockArrowUp{block}{hdr2}
    \blockArrowDown{hdr1}{block}

    \draw[->] (0.5,1) -> (4.5,1) node[midway,above] {Overflow};
\end{tikzpicture}
}

\newcommand{\BlockOverflowImageFree}{
\begin{tikzpicture}
    \blockBasicSetup{Free (Head)}{Corrupted}{Free}{Free}
    \draw (5,-1.5) rectangle node (target) {Target} ++(2.5,0.5); 

    \blockFreeEdge{hdr1}{hdr2}
    \blockArrowDown{hdr2}{target}
    \blockFreeEdge{hdr3}{hdr4}
    \blockArrowUp{block}{hdr1}

\end{tikzpicture}
}

\newcommand{\BlockOverflowUbuntuImage}{
\begin{tikzpicture}
    \blockBasicSetup{Free (Head)}{EIR Resp.}{Free}{Free}
    \draw (5,-1.5) rectangle node (target) {Target} ++(2.5,0.5); 

    \blockFreeEdge{hdr1}{hdr3}
    \blockFreeEdge{hdr3}{hdr4}
    \blockArrowDownInv{hdr2}{block}
    \blockArrowUpInv{block}{hdr1}

\end{tikzpicture}
}


\usetikzlibrary{decorations.pathreplacing}


\section{Introduction}

Bluetooth is present in a lot of privacy-sensitive applications. These include headsets that we share contacts with, smartwatches, cars, medical devices, and all kinds of \ac{IoT} products. Around \SI{4.4}{\billion} Bluetooth-enabled devices will be presumably shipped in 2020 alone, and annual device shipments are growing~\cite{2019btmarket}.

The overall zero-click attack surface is comparably large.
For example, all \emph{Apple} devices publicly expose connectable \ac{BLE} \ac{GATT} services whenever Bluetooth is enabled---even without prior pairing.
Many devices have Bluetooth enabled by default, and quite a number of them advertise their identity~\cite{milan}. Despite these identities being anonymous, an attacker might find interesting targets near airports or office buildings. Vulnerabilities are wormable, as most devices can initiate new connections.

In this work, we evaluate various attack vectors based on \ac{RCE}. We consider attacks that are either compliant with the Bluetooth 5.2 specification~\cite{bt52}, propagate into components outside of the Bluetooth chip, or brick the Bluetooth hardware. On \emph{Broadcom} combo chips, Wi-Fi and Bluetooth run on separate \ac{ARM} cores.
As they share the \SI{2.4}{\giga\hertz} antenna, they need to agree on access through \emph{coexistence mechanisms}. Using coexistence, we escalate  from Bluetooth into Wi-Fi components, block these, and then force reboot various  devices, including the \emph{iPhone 11}.

We gain Bluetooth zero-click \acs{RCE} by systematically fuzzing those parts of the \emph{Broadcom} firmware that can be reached prior to pairing.
\emph{Cypress} acquired parts of \emph{Broadcom's} Bluetooth implementation in 2016~\cite{cypressbroadcom}, and while both stacks diverged since then, they remain fuzzable and vulnerable using similar techniques.
Emulation and fuzzing provide insights into an otherwise undocumented, proprietary firmware. We provide a $C$ programming environment to interact with the firmware image that can test hypotheses on the firmware and narrow down the relevant code paths.
Our main contributions are as follows:

\begin{itemize}
  \setlength\itemsep{0em} 
\item We design and implement the emulation framework \fuzztool to execute large portions of the firmware, including injection of raw wireless frames and interaction with the host,
\item find three zero-click chip vulnerabilities, two for classic Bluetooth and one for \ac{BLE},
\item find the \emph{BlueFrag} \ac{RCE} in \emph{Android},
\item break the coexistence mechanism in Wi-Fi/Bluetooth combo chips requiring a full device reboot to restore functionality, with some devices kernel panicing,
\item uncover a design flaw in the Bluetooth 5.2 specification~\cite{bt52} that allows attackers to extract link keys including inactive connections, and
\item showcase that users cannot turn off Bluetooth as a defense on recent mobile operating systems, as the chip reset is not specified properly.
\end{itemize}

\fuzztool is publicly available on \emph{GitHub}.
The provided fuzzing examples for two \acp{CVE} find these in a matter of seconds to a few minutes.
Firmware dumps of other popular wireless systems are also good candidates to be analyzed with our solution.
We were able to confirm portability of \fuzztool by porting it to another firmware, however, we cannot present further examples due to non-disclosure agreements.

This paper is structured as follows.
\autoref{sec:overview} introduces attacks within Bluetooth stacks and clarifies the difference between the full-stack \fuzztool approach and existing wireless fuzzers.
\autoref{sec:security} showcases broader vulnerabilities and attack concepts that apply to Bluetooth chips of all manufacturers, including new exploitation techniques we found.
\autoref{sec:firmware} gives an overview of firmware and Bluetooth-specific internals. Based on this, we explain how \fuzztool works in \autoref{sec:emulation}. The identified \acp{RCE} are described in \autoref{sec:sec_analysis}.  Applicability to other firmware and vulnerability patching are discussed in \autoref{sec:discussion}. An overview of related work is given in \autoref{sec:related}. \autoref{sec:conclusion} concludes our findings.


\tikzset{>=latex}
\begin{figure*}[h]
\center
	\begin{tikzpicture}[minimum height=0.55cm, scale=0.8, every node/.style={scale=0.8}, node distance=0.7cm]

    \filldraw[darkblue, align=left, fill=white, rounded corners=4, fill=darkblue!30,](-4.5,0.5) rectangle node (fs) {} ++(14,-5); 
    \node[anchor=north,align=center, above=of fs.north,yshift=1.3cm] (fstxtb) {\textbf{Frankenstein}};

    \filldraw[gray, align=left, fill=white, rounded corners=4, fill=gray!5](3,0) rectangle node (bt) {} ++(5,-4); 
    \node[anchor=north,align=center, above=of bt.north,yshift=0.8cm] (chiptxtb) {\textbf{Virtual Bluetooth Chip}};
    \node[anchor=south,align=center, below=of bt.south,yshift=-0.8cm] (chiptxta) {Emulation \& Patching};
    \node[anchor=south,align=center, below=of bt.south,yshift=-0.8cm,xshift=5.5cm] (osint) {Pseudo Terminal};

    \filldraw[gray, align=left, fill=white, rounded corners=4, fill=gray!5, text width=3cm](3,-6) rectangle node (bt2) {\textcolor{black}{\hspace*{0.5cm}State Snapshots\\\hspace*{0.5cm}Heap Sanitizer\\\hspace*{0.5cm}PoCs}} ++(5,-2); 
    \node[anchor=north,align=center, above=of bt2.north,yshift=-0.55cm] (chiptxtb2) {\textbf{Physical Bluetooth Chip}};

    \filldraw[gray, align=left, fill=white, rounded corners=4, fill=gray!5](-3,0) rectangle node (vm) {} ++(5,-4); 
    \node[anchor=north,align=center, above=of vm.north,yshift=0.8cm] (vmtxtb) {\textbf{Virtual Modem}};
    \node[anchor=south,align=center, below=of vm.south,yshift=-0.8cm] (vmtxta) {Fuzzing Input};

    \filldraw[align=left, fill=white, rounded corners=4, fill=gray!5,](12.5,4) rectangle node (os) {} ++(5,-12); 
    \node[anchor=north,align=center, above=of os.north,yshift=4.8cm] (ostxtb) {\textbf{Operating System}};

    \filldraw[darkgreen, align=left, fill=white, rounded corners=4, fill=darkgreen!5](3,4) rectangle node (wifi) {} ++(5,-2); 
    \node[anchor=north,align=center, above=of wifi.north,yshift=-0.2cm] (wifitxtb) {\textbf{Wi-Fi Chip}};
    
    \path[->,color=darkred] (-2.75,-0.75) edge node[sloped, anchor=left, above, text width=15cm, align=left] {Extended Inquiry Response} (13,-0.75);
    \path[<-,color=darkred,dashed] (7.5,-0.25) edge node[sloped, anchor=left, above, text width=15cm, align=left] {} (13,-0.25);
    \path[->,color=darkred] (-2.75,-1.375) edge node[sloped, anchor=left, above, text width=5.5cm, align=left] {BLE PDU} (3.5,-1.375);
    \path[->,color=darkred] (-2.75,-2) edge node[sloped, anchor=left, above, text width=5.5cm, align=left] {ACL Data} (3.5,-2);
    \path[->,color=darkred] (-2.75,-3.5) edge node[sloped, anchor=left, above, text width=15cm, align=left] {LMP Fuzzing} (13,-3.5);

    \node[inner sep=0pt] (frankenstein) at (-3.5,-3.5)
    {\includegraphics[height=1.2cm]{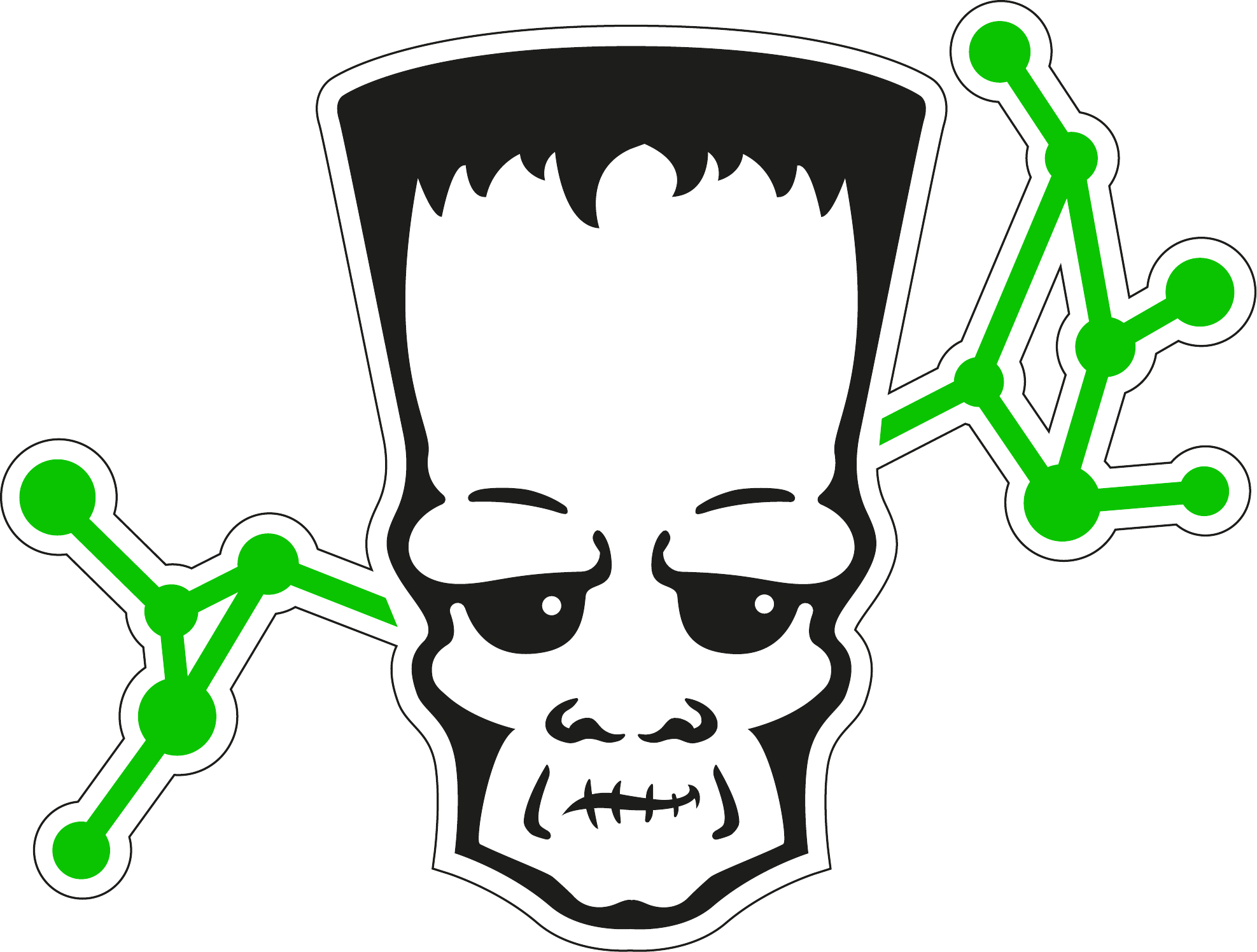}};

    \node[inner sep=0pt,align=left,anchor=west] (bam1) at (3.75,-1.375)
    {\includegraphics[height=0.7em]{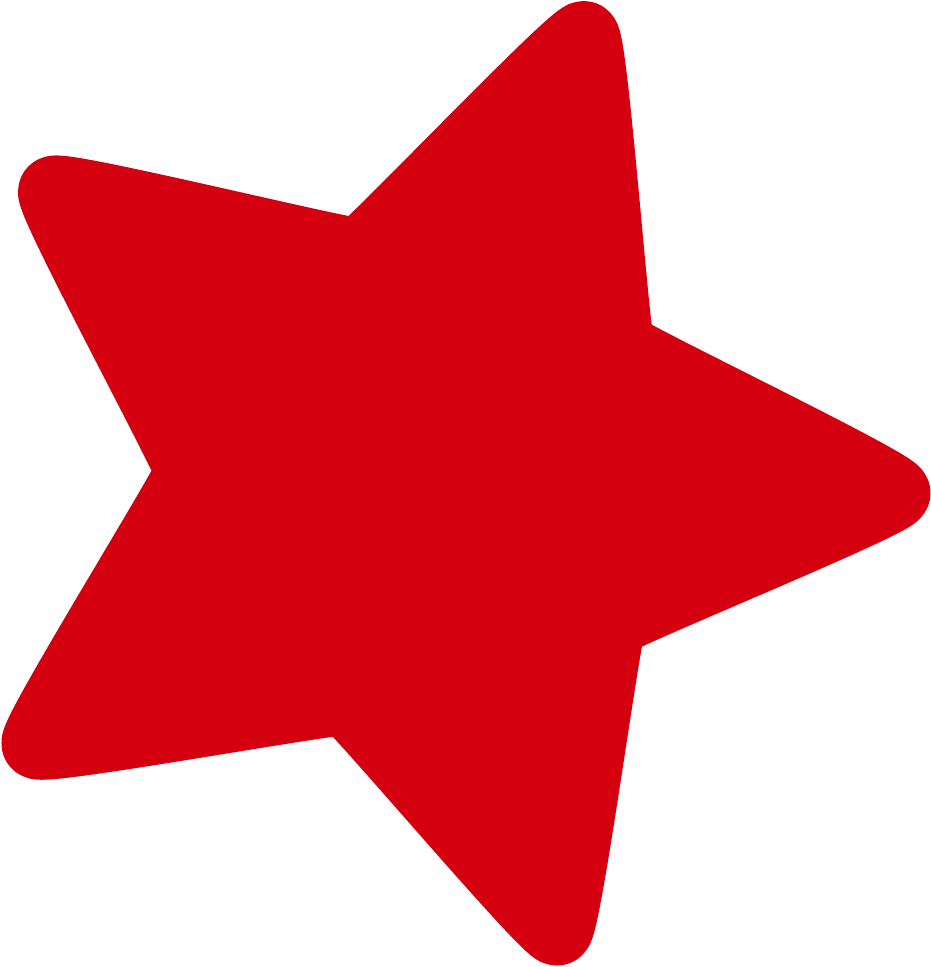}~\textcolor{darkred}{RCE  \cveb}};

    \node[inner sep=0pt,align=left,anchor=west] (bam2) at (3.75,-0.4)
    {\includegraphics[height=0.7em]{pics/bam.pdf}~\textcolor{darkred}{RCE  \cvea}};
    
    \node[inner sep=0pt,align=left,anchor=west] (bam6) at (3.75,-2)
    {\includegraphics[height=0.7em]{pics/bam.pdf}~\textcolor{darkred}{RCE  \cved}};

    \node[inner sep=0pt,align=left,anchor=west] (woop) at (5.4,-2.75)
    {\includegraphics[height=0.8em]{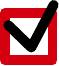}};
    \path[->,color=darkred] (5.9,-2.75) edge node[sloped, anchor=left, above, text width=5.5cm, align=left] {} (13,-2.75);
    \node[inner sep=0pt,align=left,anchor=west] (bam4) at (13.1,-2.75)
    {\includegraphics[height=0.7em]{pics/bam.pdf}~\textcolor{darkred}{Link Key Extraction}};

    \node[inner sep=0pt,align=left,anchor=west] (bam3) at (13.1,-3.5)
    {\includegraphics[height=0.7em]{pics/bam.pdf}~\textcolor{darkred}{Undetermined Issues}};
    \node[inner sep=0pt,align=left,anchor=west] (osag) at (13.5,-4) {\includegraphics[height=1.5em]{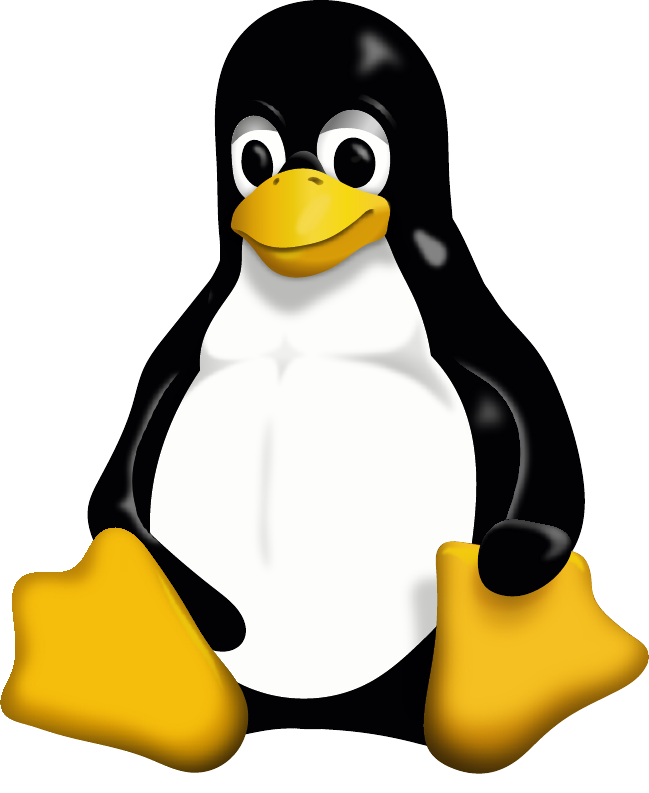}};
    
    \node[inner sep=0pt,align=left,anchor=west] (fu) at (13.1,-0.5)
    {\textcolor{darkred}{Follow-up Request}};

    \path[->,color=darkred] (5.5,0.5) edge node[anchor=left, below, text width=5cm, align=right,xshift=1.4em,yshift=0.75em] {} (5.5,2.5);
    
    \node[inner sep=0pt] (spectra) at (4.75,1.25)
    {\includegraphics[height=1cm]{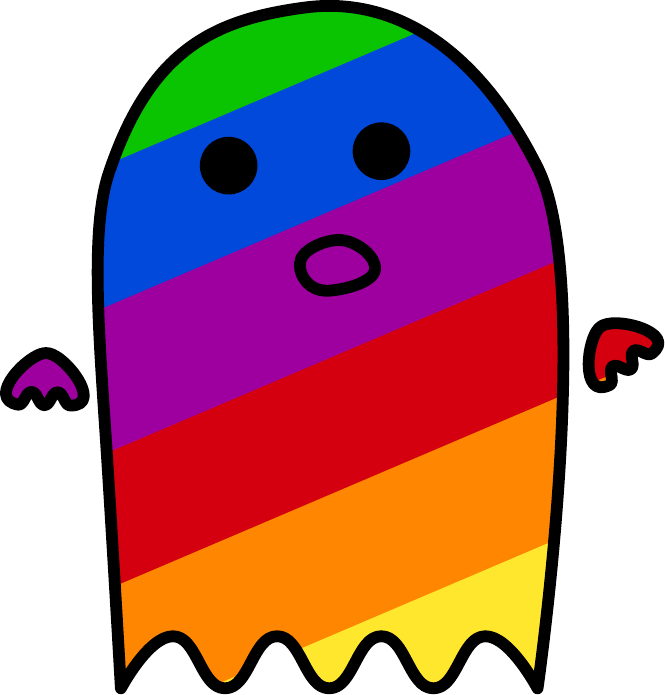}};
    
    \node[inner sep=0pt,align=right,color=darkred] (coex) at (5.5,3)
    {Coexistence DoS\\ \includegraphics[height=0.7em]{pics/bam.pdf}~\cvec};
    \path[->,color=darkred,dashed] (7.5,3) edge node[sloped, anchor=left, above, text width=2.5cm, align=left] {Driver Timeout} (13,3);
    \node[inner sep=0pt,align=left,anchor=west] (creb) at (13.1,3)
    {\textcolor{darkred}{\includegraphics[height=0.7em]{pics/bam.pdf}~Kernel Panic}};
    \node[inner sep=0pt,align=left,anchor=west] (osag) at (13.5,2.5) {\includegraphics[height=1.5em]{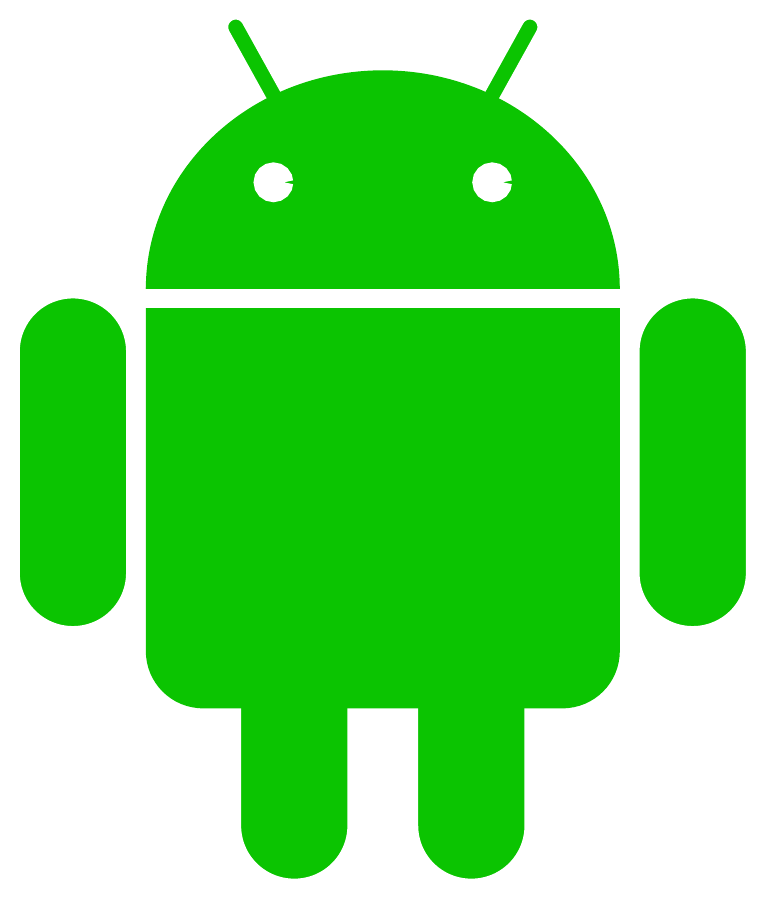}~\includegraphics[height=1.5em]{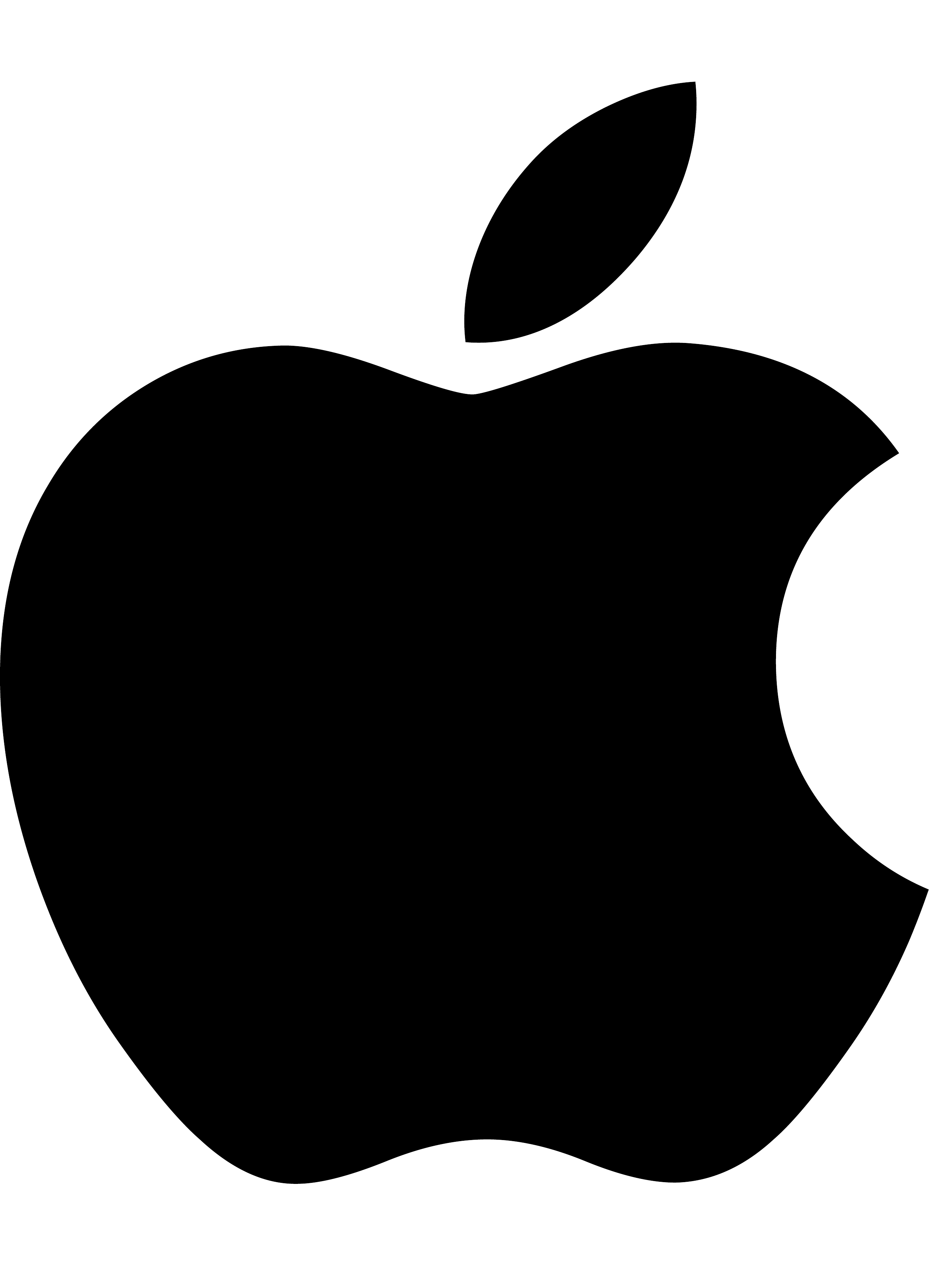}};

    \path[-,color=gray,dotted] (5.5,-4) edge node[anchor=left, below, text width=5cm, align=right,xshift=1.4em,yshift=0.75em] {} (5.5,-6);

    \path[->,color=darkred] (7.5,-7) edge node[sloped, anchor=left, above, text width=2cm, align=center] {L2CAP} (13,-7);
    \node[inner sep=0pt,align=left,anchor=west, text width=4cm] (bam5) at (13.1,-6.8)
    {\hspace*{0.7em} \textcolor{darkred}{BlueFrag RCE \newline \includegraphics[height=0.7em]{pics/bam.pdf}~\emph{CVE-2020-0022}}};
    \node[inner sep=0pt,align=left,anchor=west] (bf) at (13.5,-7.5) {\includegraphics[height=1.5em]{pics/android.pdf}};
    \node[inner sep=0pt,align=left,anchor=west] (bfl) at (9.9,-7.7) {\includegraphics[height=1cm]{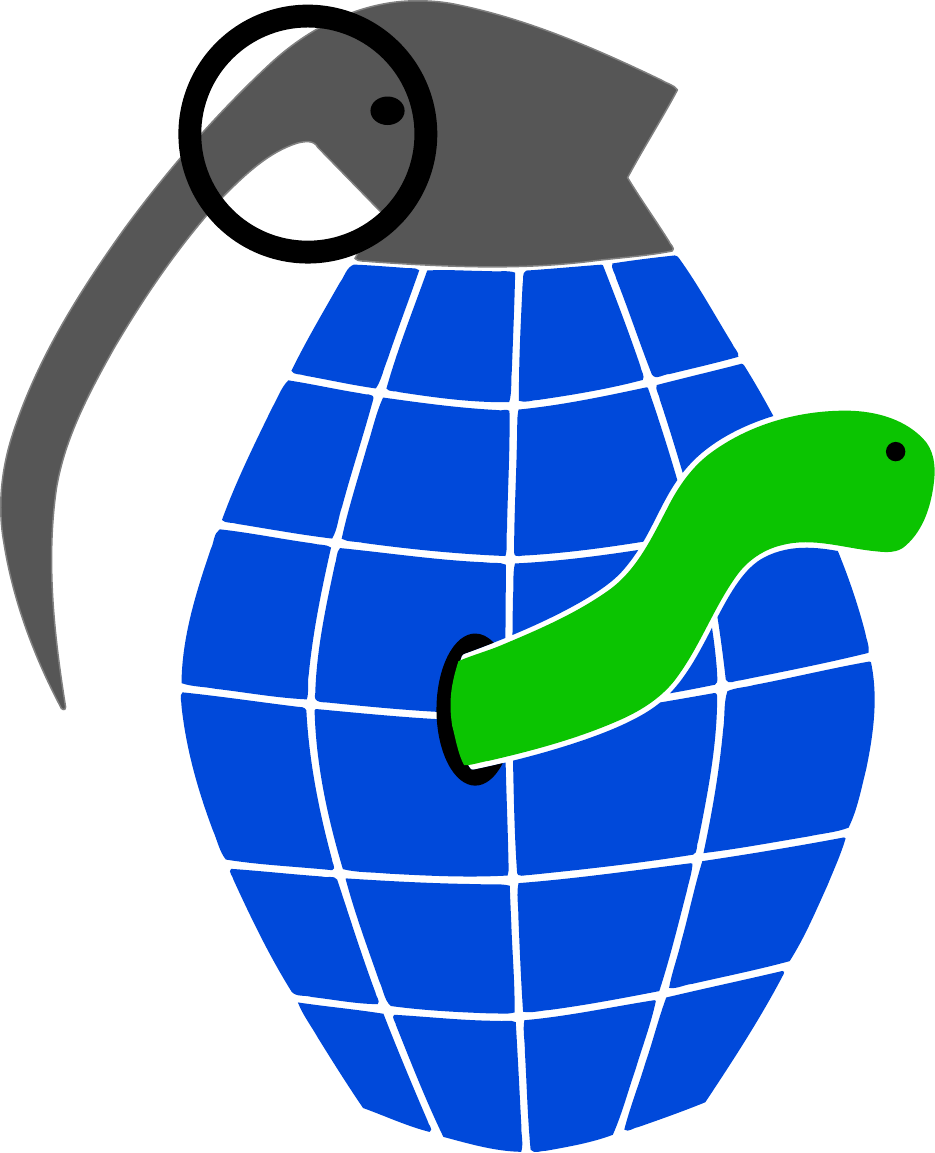}};
    
    \node[inner sep=0pt,align=left,anchor=west] (btch) at (3.3,-7.5) {\includegraphics[height=0.4cm]{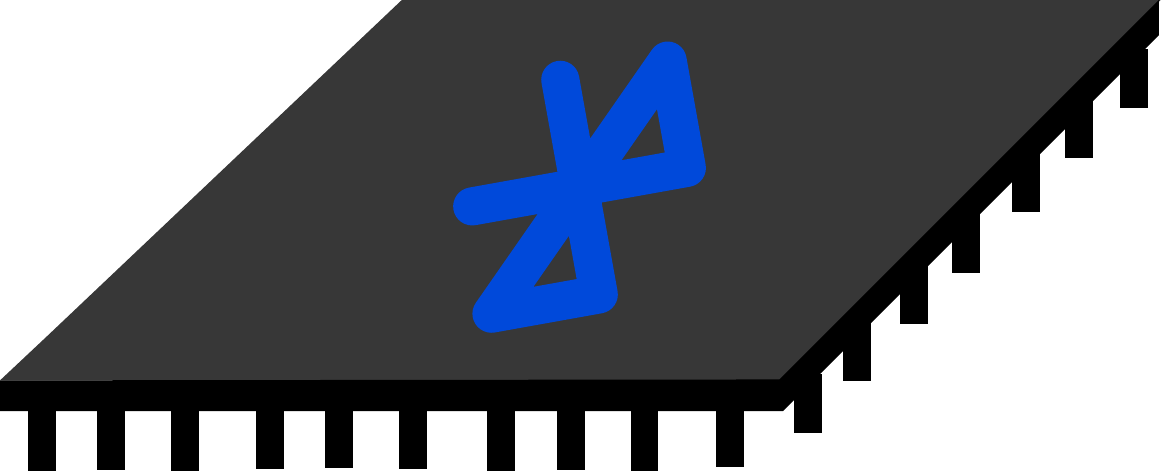}};

	\end{tikzpicture}
\caption{Bluetooth attacker model and \emph{Frankenstein} integration.} 
\vspace{-1em}
\label{fig:overview}
\end{figure*}

\section{Motivation for Frankenstein}
\label{sec:overview}
In the following, we put the motivation for \fuzztool in a broader context.
Thus, we explain the general attack paths within Bluetooth stacks in \autoref{ssec:oattacks}.
Then, we outline how \emph{Frankenstein} integrates into these stacks, how its full-stack
capability differentiates it from other fuzzers, as well as its applicability to other firmware in \autoref{ssec:ofrakenstein}.

More details about how to perform these attacks follow in \autoref{sec:security}.
A technical description of \fuzztool is provided in \autoref{sec:emulation}.
However, we recommend reading this motivation to those who are not familiar with Bluetooth and wireless fuzzing.

\subsection{Bluetooth Attack Paths}
\label{ssec:oattacks}
\autoref{fig:overview} shows the attacks uncovered with \fuzztool.
While all attacks can be launched over-the-air, their capabilities and escalation strategies differ.

\paragraph{Operating System RCE}
The most severe attacks allow direct access to the
operating system. Depending on the operating system, the Bluetooth daemon runs with limited privileges;
thus, the attacker needs to escalate further. However, on most operating systems, these
limited privileges include accessing files and contacts.

While vulnerabilities in the operating system are the most severe, they are the easiest to patch.
All they require is an operating system update, as they are hardware-independent.

\paragraph{On-Chip RCE}
The firmware running on the Bluetooth chip can be vulnerable as well. In general, it is
easier to exploit---the protection mechanisms of the \ac{RTOS} running on it and the
chip's hardware are rudimentary compared to what modern operating systems and architectures provide.
An attacker with control over the firmware can access data processed within the chip and perform
specification-compliant requests to the operating system. However, to also gain code execution on the
operating system, further vulnerabilities on the host stack are required.

Thus, despite high exploitability, full system compromise
requires additional escalation. Nonetheless, on-chip vulnerabilities are a security risk that often remains unpatched as
security fixes require patches provided by the hardware vendor that, in turn, are shipped with
an operating system update.

\paragraph{Inter-Chip Escalation}
An attack path that excludes mitigation by the operating system is inter-chip escalation.
On \emph{Broadcom} chips, Bluetooth and Wi-Fi run on two separate \ac{ARM} cores. However, to
coordinate spectrum access by means of coexistence mechanisms, they directly communicate with
each other without the operating system being involved into this. Using inter-chip escalation,
a Bluetooth \ac{RCE} can then escalate into Wi-Fi components.

Depending on the type of inter-chip escalation, this communication channel exists in hardware
and might be unpatchable. Thus, the firmware running on both cores must mitigate against this
type of attack, and the operating system drivers should take action
where possible. \\

\noindent
Within our work on \fuzztool, we uncover all these vulnerability types, as shown in \autoref{fig:overview}.
The focus of \fuzztool is to find on-chip \ac{RCE}. We show that on-chip \ac{RCE} can be used to 
break confidentiality in a specification-compliant manner by extracting the link keys used by Bluetooth encryption.
During attempts to trigger the \fuzztool vulnerabilities over-the-air, one of our \acp{PoC} triggers
\emph{BlueFrag}, an \emph{Android} operating system \ac{RCE}.
Moreover, we explore inter-chip escalations and find that we can crash the Wi-Fi firmware, which, in turn, produces
kernel panics on \emph{Android} and \emph{iOS}.

\subsection{Frankenstein}
\label{ssec:ofrakenstein}

\fuzztool creates a physical device snapshot and then emulates it in \ac{QEMU} to fuzz the full stack: over-the-air data is provided by a virtual modem, the emulated firmware implements thread and task switches to fuzz multiple handlers, and it attaches to a real \emph{Linux} host.
It utilizes \ac{QEMU} in user mode without further customizations.

\paragraph{Chip Integration and Emulation}
Firmware running on a physical chip is difficult to access, monitor, and modify.
\emph{Broadcom} provides vendor-specific commands that can be used to extract firmware from the ROM.
Moreover, the ROM can be temporarily patched with breakpoints, the so-called \emph{Patchram} mechanism.
The \emph{InternalBlue} experimentation framework enables ROM extraction and patching~\cite{mantz2019internalblue}.

The \emph{Patchram} mechanism and monitoring on the hardware itself are very limited.
Even with an over-the-air \ac{SDR} fuzzer, which would require to re-implement all the logic and
formats defined in the 3256 pages of the Bluetooth specification, analyzing the results would be infeasible.
Thus, \fuzztool fuzzes the firmware in emulation. This provides higher speed than over-the-air fuzzing
and enables coverage feedback through \ac{QEMU}.

Emulating a firmware dump comes with various challenges. These include memory map generation, chip state extraction including hardware registers, and working with only partial symbols.
The common approach to handle these challenges is to
reverse-engineer firmware in order to identify protocol parsers that pose a
potential zero-click attack surface. Then, these specific protocol handlers can be manually analyzed
or automatically fuzzed.
However, \fuzztool emulates and fuzzes the firmware as a whole---including a virtual modem
for input generation and the ability to attach it to the \emph{Linux BlueZ} Bluetooth stack.
Internally, this requires the implementation of interrupt handling and thread switches.

Instead of using the emulator for most of these tasks, \fuzztool applies these
features as \emph{C} hooks within the firmware. This enables running a selection of these hooks
on the physical chip, such as \fuzztool heap sanitizer.

\paragraph{Full-Stack Approach}
The virtual modem and the ability to interact with an operating system mean that \fuzztool triggers
realistic full-stack behavior. For example, \fuzztool generates various pairing dialogs
on an \emph{Ubuntu} desktop installation when fuzzing the \ac{LMP}. In fact, we uncover one
complex vulnerability during device scanning, where the host asks for an \ac{EIR}, and the
over-the-air reply triggers the bug. The \ac{EIR} issue is triggered by a specification-compliant message flow, meaning
that it works on both \emph{Android} and \emph{Linux} hosts.

\fuzztool is not only faster than over-the-air fuzzing, but our measurements show that it also provides significantly higher hooking performance than the state-of-the-art \emph{Unicorn} engine~\cite{unicorn}.
This speedup is required for the full-stack capability. If the fuzzer is too slow and runs into timeouts of
the operating system driver or cannot handle
interrupts and thread switches properly, attaching it to a host is impossible.

In principle, \fuzztool could also be attached to other operating systems that support running
\ac{QEMU} locally. As of June 2020, we are working on adding further operating systems.

Another, more complex application would be to replace the virtual modem with an \ac{SDR}.
While this would only be possible with a high-speed variant that supports at least \SI{80}{\mega\hertz}
bandwidth, this would result in a fully software-controllable Bluetooth stack starting at the physical layer.
Current \ac{SDR}-based Bluetooth implementations primarily support physical-layer decoding but do not provide a full stack.

\paragraph{Portability}
The main focus of this paper is the emulation of the \emph{CYW20735} Bluetooth evaluation board.
This board runs on an \ac{ARM} \emph{Cortex M4}~\cite{cypressbtoverview}.
The underling \ac{RTOS} is \emph{ThreadX}~\cite{threadx}. A more technical description of similar
platforms is provided in \autoref{ssec:applicability}.

\fuzztool requires custom hooks inside the firmware. Not accounting for Bluetooth-specific hooks,
supporting interrupts and thread switches on \ac{ARM} with \emph{ThreadX} are approximately \num{100} custom hooks.

As of June 2020, \fuzztool also partially supports the \emph{CYW20819} evaluation board as well
as the \emph{Samsung Galaxy S10/S20} firmware. For the latter, no symbols are available at all.
However, symbols are required only for the hooks. The emulation itself runs without symbols as it simply
interprets and executes binary code based on an initial state---thus,
identifying all relevant functions is sufficient.
Moreover, we used \fuzztool for a non-public project that is not a \emph{Broadcom} or \emph{Cypress}
Bluetooth chip. Although this additional project is non-public, we pushed all code changes that
enable easier integration of new projects to \emph{GitHub}.

\section{RCE-enabled Bluetooth Attacks}
\label{sec:security}

In this section, we present various novel attack scenarios enabled by on-chip \acf{RCE}.
Details on how we found and exploited on-chip \ac{RCE} in the first place are provided in \autoref{sec:sec_analysis}.
Our attacks are practical and apply to the specification, a wide variety of operating systems, or also affect the chips other than \emph{Broadcom}.

A specification-compliant attack to extract link keys is described in \autoref{ssec:link_key}. Bluetooth capabilities are typically combined with Wi-Fi within one chip, and with LTE on the same smartphone. We exploit this fact to escalate into the \mbox{Wi-Fi} chip component and cause kernel panics across various smartphone models and outline how to lower LTE performance in \autoref{ssec:coex}. An attacker might be able to brick Bluetooth chips forever, as shown in \autoref{ssec:brick}.
In general, it is hard to defend against \ac{RCE},
as turning off Bluetooth is not guaranteed to reset the chip's memory (see \autoref{ssec:disable}).

This section only discusses on-chip attacks and inter-chip escalations, as these attacks have a potential lifetime of multiple operating system major releases.
Escalations into the operating system are highly platform-dependent and rather short-lived.
Nonetheless, such escalations pose a significant threat, which has already been demonstrated as an attack for the \emph{Broadcom} Wi-Fi implementation~\cite{2017:googleprojectzero,2017:artenstein,quarkslab2019}.
Since the \emph{iPhone XS}, the \ac{HCI} is attached via \ac{PCIe}, exposing similar escalation targets.

\begin{table*}[!t]
\renewcommand{\arraystretch}{1.3}
\caption{Exploiting Wi-Fi through Bluetooth coexistence on combo chips (\cvec).}
\label{tab:coex}
\centering
\scriptsize
\begin{tabular}{|l|p{2.1cm}|p{1.7cm}|r|p{1.2cm}|p{0.8cm}|p{6.2cm}|}
\hline
\textbf{Chip} & \textbf{Device} & \textbf{OS} & \textbf{Build Date} & \textbf{Address} & \textbf{Value} & \textbf{Effect}\\
\hline
BCM4335C0 & Nexus 5 & Android 6.0.1 & Dec 11 2012 & \texttt{0x650440}, \texttt{0x650600} & \texttt{0x00} & Disconnects from \SI{2.4}{\giga\hertz} and \SI{5}{\giga\hertz} Wi-Fi, Wi-Fi can be reconnected. \\
BCM4345B0 & iPhone 6 & iOS 12.4 & Jul 15 2013 & \texttt{0x650000}--\texttt{0x6507ff} &  & Disables \SI{2.4}{\giga\hertz} Wi-Fi until restarting Bluetooth. \\
\hline
BCM4345C0 & Raspberry Pi 3+/4 & Raspbian Buster & Aug 19 2014 & \texttt{0x650000}--\texttt{0x6507ff} & Random & Full and partial Wi-Fi crashes, including \ac{SDIO}, ability to scan for Wi-Fis, speed reduction. Reboot required to restore functionality.\\
BCM4358A3 & Nexus 6P & Android 7.1.2 & Oct 23 2014  & \texttt{0x650000}--\texttt{0x6507ff} &  & Disables all Wi-Fi until restarting Bluetooth. \\
BCM4358A3 & Samsung Galaxy S6 & Lineage OS 14.1 & Oct 23 2014  & \texttt{0x650000}--\texttt{0x6507ff} &  & Disables all Wi-Fi until restarting Bluetooth. \\
\hline
BCM4345C1 & iPhone SE & iOS 12.4--13.3.1 & Jan 27 2015 & \texttt{0x650200} & \texttt{0xff} & Kernel panic, resulting in a reboot. \\
BCM4355C0 & iPhone 7 & iOS 12.4--13.3.1 & Sep 14 2015 & \texttt{0x650200} & & Kernel panic, resulting in a reboot. \\
BCM4347B0 & Samsung Galaxy S8 & Android 8.0.0 & Jun 3 2016 & \texttt{0x650200} &  & Disables \SI{2.4}{\giga\hertz} and \SI{5}{\giga\hertz} Wi-Fi, kernel panic and reboot when re-enabling Wi-Fi. \\
BCM4347B0 & Samsung Galaxy S8 & LineageOS 16.0 & Jun 3 2016 & \texttt{0x650200} &  & Temporarily disables \SI{2.4}{\giga\hertz} and \SI{5}{\giga\hertz} Wi-Fi, freezes system for a couple of seconds when re-enabling Wi-Fi. \\
BCM4347B1 & iPhone 8/X/XR & iOS 12.4--13.3.1 & Oct 11 2016 & \texttt{0x650200} &  & Kernel panic, resulting in a reboot. \\
BCM4375B1 & Samsung Galaxy\newline S10/S10e/S10+ & Android 9 & Apr 13 2018 & \texttt{0x650200} &  & Disables \SI{2.4}{\giga\hertz} and \SI{5}{\giga\hertz} Wi-Fi. Reboot required to re-enable Wi-Fi.\\
BCM4377B3 & MacBook Pro/Air\newline 2019--2020 & macOS \newline 10.15.1--10.15.5 & Feb 28 2018 & \texttt{0x650400} &  & Kernel panic, resulting in a reboot.\\
BCM4378B1 & iPhone 11 & iOS 13.3 & Oct 25 2018 & \texttt{0x650400} &  & Kernel panic, resulting in a reboot.\\
\hline
\end{tabular}
\vspace{-1em} 
\end{table*}

\subsection{Link Key Extraction}
\label{ssec:link_key}

During initial pairing between two devices, a link key is negotiated. It will ensure the security of all follow-up connections between the two paired devices.
If the link key of a user's headset leaks, an attacker can listen to calls and access the user's phone book. Paired keyboards and mice can generate arbitrary input or be eavesdropped. \emph{Smart Lock}, introduced in \emph{Android 5} and still present in \emph{Android 10}~\cite{smartlock}, enables users to unlock their smartphone with nearby paired devices.

A Bluetooth implementation can either hold the link keys within the controller or on the host.
The \emph{Broadcom} chip has no permanent storage except the ROM. Thus, the host
stores link keys for all connections. According to the Bluetooth specification~\cite[p. 1948]{bt52}, the controller can ask the host for a link key associated with a \ac{MAC} address.
The host will send back different message types depending on whether it has a link key for a requested MAC address.
This separation into two message types simplifies exploitation. For example, an attacker can hook the reply function  
 inside the \emph{Broadcom} chip to copy the link key to the global device name variable.
Reusing existing firmware functions makes this patch require around \SI{128}{\byte} in practice~\cite{classen2019recon}.

The ability of the controller to request any encryption key differs a lot from other wireless standards. It is very specific to Bluetooth, because the
simple pairing concept of \ac{TOFU} also means that there is no additional verification by certificates or other external dependencies.
In contrast to existing attacks on pairing and key negotiation~\cite{knob,2018:biham}, our link key extraction does \emph{not} require an active \ac{MITM} setup, but \ac{RCE}.

Our tests on real devices showed that even the link key for inactive connections could be requested.
As a link key extraction countermeasure, the host should only return link keys if proper \ac{HCI} messages were exchanged previously.
For example, \emph{BTstack} only copies the link key if it has an active connection~\cite{btstack}.
Moreover, the stack should introduce a short delay in link key request replies to prevent \ac{MAC} address brute-force attacks.

However, this attack can only be made harder, but cannot be prevented completely while keeping the host's implementation Bluetooth specification-compliant.
As any proper mitigation would break compatibility with the current specification, including the whole \ac{TOFU} concept that enables Bluetooth pairing without
certificate checks, we did not report this issue to the Bluetooth SIG but only to the vendors.
In general, vendors are aware of this---\emph{Apple} even designed \emph{MagicPairing} to secure pairing of their proprietary Bluetooth peripherals and
integrate them into \emph{iCloud}~\cite{magicpairing}.

\subsection{Inter-Chip Escalation (\cvec)}
\label{ssec:coex}

In the following, we analyze possibilities to escalate from Bluetooth into further wireless components.
This is possible because Wi-Fi and Bluetooth are combined in the same chip, and reside with LTE on the same smartphone. 
On \emph{Broadcom} Wi-Fi/Bluetooth combo chips, each protocol runs on a separate \ac{ARM} core, but they share parts of the transceiver. They have a common interface to communicate their needs, which we exploit to shut down Wi-Fi persistently. 
The operating system cannot prevent this type of inter-chip escalation.

Coexistence between Bluetooth and Wi-Fi is usually realized by applying an
\ac{AFH} channel map~\cite[p. 289]{bt52},
which can blacklist overlapping \SI{2.4}{\giga\hertz} channels.
Vendors can implement proprietary coexistence additions for better performance~\cite[p. 290]{bt52}.
Simply blacklisting channels is not sufficient on \emph{Broadcom} Bluetooth combo chips---they add their own \ac{ECI} protocol. 
\ac{ECI} optimizes priorities for different types of Wi-Fi and Bluetooth packets. Each protocol stack collaboratively waits for the other, depending on the scenario.

Our practical tests disabling coexistence confirm that \emph{Broadcom} combo chip performance highly depends on it. When streaming a video and simultaneously listening to it with Bluetooth headphones, the video stutters while the sound is playing for a few seconds, and then the sound stops while the video continues buffering.
This means, as a countermeasure against attacks on coexistence, \emph{Broadcom} cannot simply disable it. \SI{2.4}{\giga\hertz} Wi-Fi and Bluetooth would block each other significantly, even without any attacker being present.

Coexistence implementations vary a lot between chips. 
While there are different implementations, firmware compiled between 2012 and 2018\footnote{Chips require at least a year to appear in the wild, and this is the newest firmware we had access to as of June 2020. The latest \emph{iPhone SE2}, \emph{MacBook Pro 2020}, and \emph{Samsung Galaxy S20} all use firmware dating back to 2018.} map coexistence registers to the same memory area.
We crash or practically disable Wi-Fi by writing to those registers via Bluetooth, as listed in \autoref{tab:coex}. Often, it is impossible to re-enable Wi-Fi, and the device needs to be rebooted to restore functionality. The \emph{Samsung Galaxy S8} stock ROM tries to re-enable \mbox{Wi-Fi} five times until rebooting with a soft kernel panic. 
When installing a \emph{LineageOS 16.0} unofficial nightly build from August 30 2019, and performing the same attack on the \emph{Samsung Galaxy S8}, the log shows errors related to \texttt{WifiHAL}. While \emph{LineageOS 16.0} does not reboot, the screen is still freezing for a couple of seconds, then turns off and leaves the user at the lock screen. We also observed a kernel panic on the \emph{iPhone SE}, \emph{7}, \emph{8/X/XR}, and \emph{11} related to a kernel mutex and \texttt{AppleBCMWLANBusInterfacePCIe}.

In general, coexistence can also be disabled in other ways, such as ignoring callbacks with channel blacklistings or packet transmission requests. The attack also works the other way round---we produced a Wi-Fi firmware that never allows Bluetooth to transmit on the \mbox{\emph{Nexus 5}} with \emph{Nexmon}~\cite{schulz2018}.

Coexistence for shared or co-located antennas is also an issue across vendors.
Various frequency bands of technologies used within one device are likely to interfere with Bluetooth, including LTE bands 40 and 7 uplink close to the \SI{2.4}{\giga\hertz} band. 
In addition to those direct neighbors, harmonics can also interfere.
 Advanced measurement setups in shielded chambers allow measuring the exact interference within a given device~\cite{coex_appnote}.

Vendor-independent solutions enable coexistence between Bluetooth, Wi-Fi, and LTE chips.
The Bluetooth specification outlines a generic \acf{MWS}  scheme for coexistence with both LTE and Wi-Fi~\cite[p. 3227ff]{bt52}.
\emph{Broadcom} implements all \ac{MWS} \ac{HCI} commands the specification proposes, along with vendor-specific additions. 
This enables LTE coexistence with chips of different manufacturers, such as \emph{Intel} or \emph{Qualcomm}. 
Since \ac{MWS} coexistence is coupled less tightly to the hardware than \ac{ECI}, we assume that tampering with \ac{MWS} commands only leads to performance degradation, but no kernel panics. Performance issues highly depend on the chip-internal implementations as well as physical aspects such as the frequency and antenna location.

Indeed, \ac{MWS} is used on \emph{iPhones}. The \texttt{WirelessRadio} 
 \texttt{Managerd} manages coexistence between LTE, Bluetooth, and Wi-Fi. 
We can observe \ac{MWS} messages on various \emph{iPhone} models. In contrast, we could not see any \ac{MWS} messages on the \emph{Samsung Galaxy S8} and \emph{S10e}.

\subsection{Bricking Hardware}
\label{ssec:brick}

At first sight, \emph{Broadcom's} memory layout seems unbrickable.
Firmware is stored in ROM, and patches are temporarily applied in Patchram. After a hard reset, all changes are gone.
Though, there is a \ac{NVRAM} section that should only be written during manufacturing. It contains a per-device configuration
like the \ac{MAC} address and crystal trimming information.

The \emph{WICED Studio} documentation warns users
about writing to \ac{NVRAM} slots below \texttt{0x200}. 
The \emph{WICED} \ac{HAL} only accepts higher slots.
An attacker can skip this \ac{HAL} safety mechanism and directly call the \path{nvram_write} function. 
We did not want to brick our Bluetooth devices, yet our experiments writing to \ac{NVRAM}  bricked one \emph{Broadcom} Wi-Fi evaluation board.

While it might still be possible to recover a device to a non-bricked state, this requires system-level access to the Bluetooth controller. On a smartphone, this implies either a patch issued by the manufacturer or the user taking control over the device to unbrick Bluetooth. The latter is an obstacle on \emph{iPhones}, which require to be jailbroken for this, and \emph{Samsung} devices, which flip the \emph{Knox} bit once rooted.

\subsection{Ineffective Defense: Disabling Bluetooth}
\label{ssec:disable}

On recent mobile operating systems, turning off Bluetooth via the advanced settings menu will not turn the chip off.
This is counter-intuitive because active connections to other devices are lost.
We test \ac{RCE} persistence by checking if memory is reset and timers continue running. 
The underlying flaw is in the Bluetooth specification, which allows a soft reset.

\subsubsection{\ac{HCI} Reset}
According to the Bluetooth 5.2 specification, the \path{HCI_Reset} command will not necessarily perform a hardware reset~\cite[p. 2077]{bt52}. On the \emph{CYW20735} evaluation board, only some timers, current connections, link manager queues, and similar information are reset. No full hardware reset is performed.

\subsubsection{Testing Chip Hard Reset}
We analyze if a device was appropriately reset. On \emph{Broadcom} and \emph{Cypress} firmware, a \texttt{bootcheck} memory area
is written during a hard \texttt{\_\_reset} of any device under test. We insert custom values into this area. If they stay persistent, we know that no hard reset took place. This approach excludes that memory is persistent due to cold boot effects~\cite{ullrich2019neato}. Moreover, timer registers can be used to confirm the hardware state.
We issue \path{HCI_Reset} commands on chips ranging from 2012 to 2018. Indeed, the \texttt{bootcheck} memory area is never reset.

\subsubsection{iOS Devices}

On \emph{iOS 12} and \emph{13} devices, including \emph{iOS 13.5}, the Bluetooth chip is neither hard reset when Bluetooth is disabled nor in flight mode. Under some circumstances, like a firmware crash, a hard reset can happen. When Bluetooth is disabled via the settings menu, we can still connect to other devices when issuing commands on the chip.
Executing commands on the chip and getting \ac{HCI} events passed to the host for processing connection establishments requires \texttt{btwake} to be active. We believe this to not be a showstopper when facing \ac{RCE}, since it is implemented as interrupt on the firmware and can be reconfigured. Communication with the host is not necessarily required when adding functions inside the firmware to handle over-the-air requests.

While Bluetooth is enabled on an \emph{iPhone}, it can be found using \ac{BLE} device scanning. The \ac{MAC} address is randomized, but an attacker can connect and request the firmware version.
\ac{BLE} advertisements contained a device name in \emph{iOS 12}~\cite{milan}, which has been fixed in \emph{iOS 13}. 
However, this anonymity does not stop attackers, as Bluetooth requires proximate targets either way. Moreover, Bluetooth has become an even more integral part of \emph{iOS 13} due to features like \emph{Find My}~\cite{findmy}.

\subsubsection{Android Devices}

In contrast to \emph{iOS}, \emph{Android 8} and \emph{9} on a \emph{Samsung Galaxy S8} as well as \emph{Android 9} and \emph{10} on a \emph{Samsung Galaxy S10e} will disable and hard reset Bluetooth in flight mode. However, when not in flight mode, the Bluetooth chip will not be reset by turning off Bluetooth. The latest version we tested is \emph{Android 10} on the March 2020 patch level.
This behavior does not change when disabling location services. Whenever a user turns off Bluetooth, only \ac{BLE} and classic scanning for devices are disabled. No \path{HCI_Reset} is issued. It is still possible to connect to other devices.

\emph{Android 6} on a \emph{Nexus 5} resets memory contents and also reloads the firmware patch file with each Bluetooth restart.


\section{Proprietary Firmware Internals}
\label{sec:firmware}

Understanding firmware internals is essential to master emulation and find on-chip \ac{RCE} vulnerabilities.
\autoref{fig:firmware} depicts firmware internals, which we explain top-down in the following.
The details described in this section were discovered and analyzed using the emulation techniques described later in \autoref{sec:emulation}. 
Our analysis is based on the \emph{Cypress CYW20735} evaluation board and its firmware~\cite{cypress_cyw20735}, which was shipped with partial symbols in the \emph{WICED Studio 6.2} toolsuite~\cite{mantz2019internalblue}.
For this firmware, no public documentation or source code is available.

\subsection{Interaction Between Host and Controller}
\label{ssec:uarthostcontroller}
In Bluetooth terminology, the host is the operating system, and the controller is the chip running the firmware.
A host communicates with the controller using the \ac{HCI}. 
In the case of the \emph{CYW20735} chip, \ac{HCI} is sent via \ac{UART} to the host. 
Data is sent via the same interface using \ac{ACL} and \ac{SCO} packets.
Data does not require any interpretation by the \ac{LM}. 

\subsection{ThreadX}
\label{sec:block}

The firmware is based on \emph{ThreadX}, a \ac{RTOS} optimized for embedded devices~\cite{threadx}.
\emph{ThreadX} implements threads, events, queues, semaphores, and dynamic memory.
The firmware uses several threads, such as the \ac{LM}, \ac{UART} state machine, and a special idle thread.
Each thread implements a main loop, waiting for events to be processed.
When an event for a waiting thread is created, a context switch is performed.
Those events are mainly used for inter-thread communication, i.e., pass an \ac{HCI} packet from the \ac{LM} to the \ac{UART} state machine.
If all events are processed, the firmware enters an idle state.
At this point, new events are only generated by interrupts, such as \ac{UART}, the \ac{BCS}, or timers.

\newcommand{\tikzantenna}[1]{
    \draw [-] ($(#1)+(0.2, 0.4)$) -> (#1);
    \draw [-] ($(#1)+(-0.2, 0.4)$) -> (#1);
    \draw [-] ($(#1)+(-0.2, 0.4)$) -> ($(#1)+(0.2, 0.4)$);
    \draw [-] ($(#1)+(0, 0.4)$) -> (#1);

}
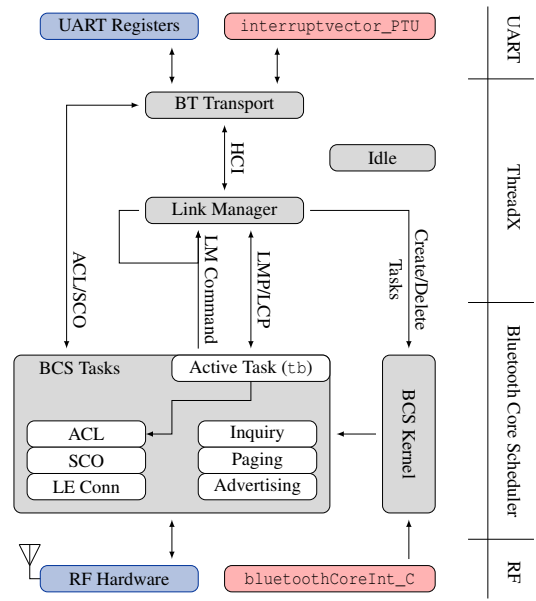
\begin{figure}[!b]
\centering
\vspace{-0.5em} 
\scalebox{0.7}{
\begin{tikzpicture}
    \node[align=left] at (9, 1.75) (legendStart) {}; \node[align=left] at (9,-9.75) (legendStop) {};
    \draw [-] (legendStart) -> (legendStop);
    \node[align=left] at (8.5, 0.25) (legendStart) {}; \node[align=left] at (10, 0.25) (legendStop) {};
    \draw [-] (legendStart) -> (legendStop);
    \node[align=left] at (8.5, -4) (legendStart) {}; \node[align=left] at (10, -4) (legendStop) {};
    \draw [-] (legendStart) -> (legendStop);
    \node[align=left] at (8.5, -8.5) (legendStart) {}; \node[align=left] at (10, -8.5) (legendStop) {};
    \draw [-] (legendStart) -> (legendStop);
    \node[align=right, rotate=270] at (9.5,1) (patch1) {UART};
    \node[align=right, rotate=270] at (9.5,-2) (patch1) {ThreadX};
    \node[align=right, rotate=270] at (9.5,-6.25) (patch1) {Bluetooth Core Scheduler};
    \node[align=right, rotate=270] at (9.5,-9.125) (patch1) {RF};

    \filldraw[darkblue, align=left, fill=white, rounded corners=4, fill=darkblue!30,](0.5,1.5) rectangle node (host) {\textcolor{black}{UART Registers}} ++(3,-0.5); 
    \node[align=left] at (1.5,1.5) (uartb) {};
    \filldraw[align=left, rounded corners=4, fill=red!30,](4,1.5) rectangle node (ptu) {\texttt{interruptvector\_PTU}} ++(4,-0.5); 
    \node[align=left] at (7.5,1.5) (ptur) {};

    \node[align=left] at (4,-0.5) (bttb) {}; \node[align=left] at (4,0) (bttt) {};
    \node[align=left] at (2.5,-0.25) (bttl) {}; \node[align=left] at (5.5,-0.25) (bttr) {};
    \filldraw[align=left, fill=gray!30, rounded corners=4](2.5,0) rectangle node (btt) {BT Transport} ++(3,-0.5); 
    \filldraw[align=left, fill=gray!30, rounded corners=4](6,-1) rectangle node () {Idle} ++(2,-0.5); 

    \node[align=left] at (4,-2) (lmt) {};
    \node[align=left] at (2.5,-2.25) (lml) {}; \node[align=left] at (5.5,-2.25) (lmr) {};
    \node[align=left] at (3.5,-2.5) (lmbl) {}; \node[align=left] at (4.5,-2.5) (lmbr) {};
    \filldraw[align=left, fill=gray!30, rounded corners=4](2.5,-2) rectangle node (lm) {Link Manager} ++(3,-0.5); 

    \filldraw[align=left, fill=gray!30, rounded corners=4](0,-5) rectangle node (bcs) {} ++(6,-3); 
    \node[align=left] at (1.25,-5.25) (cwbcstasks) {BCS Tasks}; 
    \filldraw[align=left, fill=white, rounded corners=4](3,-5 ) rectangle node (bcsactive) {Active Task (\texttt{tb})} ++(3,-0.5); 
    \node[align=left] at (3,-8) (bcstaskb) {}; \node[align=left] at (6,-6.5) (bcstaskr) {};

    \filldraw[align=left, fill=white, rounded corners=4](0.25,-6.25 ) rectangle node (aa) {ACL} ++(2.25,-0.5); 
    \filldraw[align=left, fill=white, rounded corners=4](0.25,-6.75) rectangle node (aa) {SCO} ++(2.25,-0.5); 
    \filldraw[align=left, fill=white, rounded corners=4](0.25,-7.25 ) rectangle node (aa) {LE Conn} ++(2.25,-0.5); 

    \filldraw[align=left, fill=white, rounded corners=4](3.5,-6.25 ) rectangle node (aa) {Inquiry} ++(2.25,-0.5); 
    \filldraw[align=left, fill=white, rounded corners=4](3.5,-6.75) rectangle node (aa) {Paging} ++(2.25,-0.5); 
    \filldraw[align=left, fill=white, rounded corners=4](3.5,-7.25 ) rectangle node (aa) {Advertising} ++(2.25,-0.5); 

    \node[align=left] at (7,-6.5) (bcsl) {};
    \node[align=left] at (7.5,-5) (bcst) {}; \node[align=left] at (7.5,-8) (bcsb) {};
    \node[align=left] at (3.5,-5) (bcstl) {}; \node[align=left] at (4.5,-5) (bcstr) {};
    \node[align=left] at (1,-5) (bcstll) {};
    \filldraw[align=left, fill=gray!30, rounded corners=4](7,-5) rectangle node[rotate=270] (bcs) {BCS Kernel} ++(1,-3); 

    \node[align=left] at (3,-9) (rft) {}; \node[align=left] at (0.75,-9.25) (rfl) {};
    \draw [<->] (bcstaskb) -> (rft);
    \tikzantenna{0.3,-9}
    \draw [-] (0.3,-9) |- (rfl);
    \filldraw[darkblue, align=left, fill=darkblue!30, rounded corners=4](0.5,-9) rectangle node (rf) {\textcolor{black}{RF Hardware}} ++(3,-0.5); 

    \node[align=left] at (7.5,-9) (btcoreint) {};
    \filldraw[align=left, fill=red!30, rounded corners=4](4,-9) rectangle node (mut) {\texttt{bluetoothCoreInt\_C}} ++(4,-0.5); 

    \draw [->] (lmr)  -| node [sloped, above, pos=.75] {Create/Delete} node [sloped, below, pos=.75] {Tasks} (bcst);
    \draw [->] (btcoreint) -> (bcsb);
    \node [rotate=270] at (7.75, -8.5) {};
    \draw [->] (lml)  -| node [above] {} ++(-0.5,0) |- node [above] {} ++(0,-1) -| (lmbl);
    \draw [->] (bcstl) -> (lmbl);
    \node [rotate=270] at (3.75,-3.75) {LM Command};
    \draw [<->] (bcstr) -> (lmbr);
    \node [rotate=270] at (4.75,-3.75) {LMP/LCP};

    \draw [<->] (bcstll) node [above] {} |- (bttl);
    \node [rotate=270] at (1.25,-3.75) {ACL/SCO};

    \draw [<->] (bttb) -> (lmt);
    \node [rotate=270] at (4.25,-1.25) {HCI};

    \node [rotate=270] at (7.75, 0.5) {};

    \draw [<->] (3,0.85) -- (3,0.15);
    \draw [<->] (5,0.85) -- (5,0.15);

    \draw [->] (4.5,-5.5) node [above] {} |- ++(0,-0.3625) -|  (3, -6.5) |- (2.5, -6.5);

    \draw [->] (bcsl) node [above] {} -> (bcstaskr);

\end{tikzpicture}}
\vspace{-0.5em} 
\caption{\emph{Broadcom/Cypress} Bluetooth firmware internals.}
\label{fig:firmware}
\end{figure}

\subsection{Bluetooth Core Scheduler}
The \ac{BCS} is a separate component, handling time-critical Bluetooth events.
The interrupt handler \path{bluetoothCoreInt_C} calls it every \SI{312.5}{\micro \second}. This timing is the smallest unit of the Bluetooth clock and corresponds to \nicefrac{1}{2} slot length~\cite[p. 415]{bt52}.
The \ac{BCS} kernel holds a pool of various tasks, whereas only a single task can be active at any point in time.
Tasks implement \ac{ACL}, device inquiry, paging, and more.
They directly access the hardware packet buffer and registers holding packet information.
For classic Bluetooth, which supports higher throughput rates than \ac{BLE}, the packet buffer is mapped into RAM using a \ac{DMA} mechanism.
On reception, the packet is copied to dynamic memory and handed to the corresponding thread.


\section{The Frankenstein Framework}
\label{sec:emulation}

We call our firmware emulation framework \fuzztool because it modifies a firmware image to bring it back to life in a different environment.
Snapshots of the state of the physical hardware during normal operation can be integrated and ported to the emulated environment.
In the following, we showcase the capabilities of our approach on the \emph{CYW20735} Bluetooth controller. However, other firmwares are also supported.
The emulated virtual Bluetooth chip can even be attached to a sophisticated operating system like \emph{Linux}, but in principle also to other operating systems that support \ac{UART} Bluetooth, such as \emph{macOS}.
All steps to revive the \emph{CYW20735} firmware are explained in the following.

\subsection{Bringing Firmware Images Back to Life}

Emulation either requires firmware initialization or a clean memory snapshot containing all registers.
Memory snapshots simplify the process for complex firmware.
Initially, it might be undocumented how memory is mapped. Thus, \fuzztool comes with
a \path{map_memory} hook that overwrites the \ac{ARM} memory fault handler and sweeps through
the whole address range.
Once the memory map is known, a snapshot of the memory is
obtained from a physical device by executing an \path{xmit_state} hook, which can be placed within arbitrary functions.
The \path{xmit_state} hook pauses interrupts and disables the watchdog while copying all memory via \ac{HCI}, which takes several minutes.
Since snapshot hooks are placed within functions, the snapshot state is comparably deterministic.
For example, snapshots can be taken while the chip has an active connection 
within a selected protocol handler.

\begin{figure}[!t]
\centering
\scalebox{0.7}{
\begin{tikzpicture}
    \draw[align=left,draw=none](0,1.5) rectangle node (dev) {\textbf{Physical Chip}} ++(2.0,0.5); 
    \node at (2.5, -0.5) (firmware) {};
    \filldraw[gray,align=left, fill=gray!5, rounded corners=4](-0.5,1.5) rectangle node (buff1) {} ++(3,-3.5); 
    \filldraw[align=left, fill=white](0,1) rectangle node (buff1) {Firmware} ++(2.0,0.5); 
    \node at (2.5, -0.5) (firmware) {};
    \filldraw[align=left, fill=white](0,-0  ) rectangle node (buff1) {\texttt{ROM}} ++(2.0,0.5); 
    \filldraw[align=left, fill=white](0,-0.5) rectangle node (buff1) {\texttt{RAM}} ++(2.0,0.5); 
    \filldraw[align=left, fill=white](0,-1.0) rectangle node (buff1) {\texttt{MMIO1}} ++(2.0,0.5); 
    \filldraw[align=left, fill=white](0,-1.5) rectangle node (buff1) {\texttt{MMIO2}} ++(2.0,0.5); 
    \node[inner sep=0pt,align=left,anchor=west] (btch) at (-0.2,-1.6) {\includegraphics[height=0.4cm]{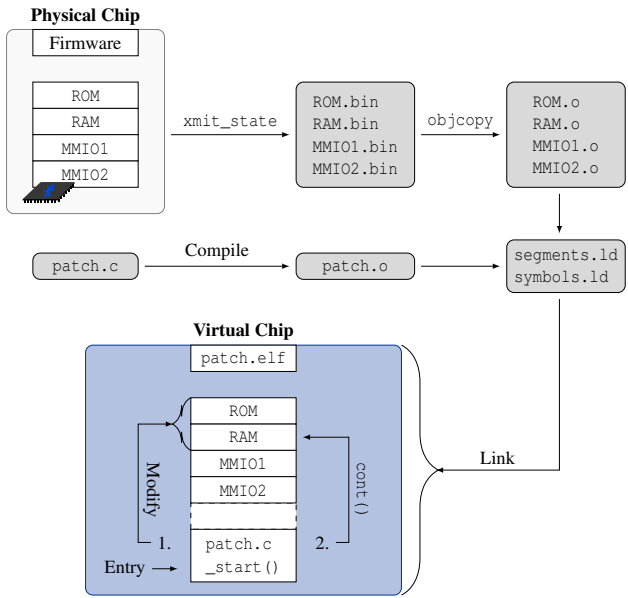}};

    \node[align=left] at (5,-0.5) (dumpBinl) {}; \node[align=left] at (7.25,-0.5) (dumpBinr) {};
    \filldraw[align=left, fill=gray!30, rounded corners=4] (5,-1.5) rectangle node (dumpBin)
        {\texttt{ROM.bin}\\\texttt{RAM.bin}\\\texttt{MMIO1.bin}\\\texttt{MMIO2.bin}} ++(2.25,2.0);

    \node[align=left] at ( 9,-0.5) (dumpOl) {}; \node[align=left] at (10,-1.5) (dumpOb) {};
    \filldraw[align=left, fill=gray!30, rounded corners=4] (9,-1.5) rectangle node (dumpO)
        {\texttt{ROM.o}\\\texttt{RAM.o}\\\texttt{MMIO1.o}\\\texttt{MMIO2.o}} ++(2.25,2.0);

    \node[align=left] at (9,-3) (ldScriptsl) {};
    \node[align=left] at (10,-2.5) (ldScriptst) {}; \node[align=left] at (10,-3.5) (ldScriptsb) {};
    \filldraw[align=left, fill=gray!30, rounded corners=4] (9,-3.5) rectangle node (ldScripts)
        {\texttt{segments.ld}\\\texttt{symbols.ld}} ++(2.25,1.0);

    \node[align=left] at (2,-3) (patchCr) {};
    \filldraw[align=left, fill=gray!30, rounded corners=4] (0,-3.25) rectangle node (patchC)
        {\texttt{patch.c}} ++(2.,0.5);

    \node[align=left] at (5,-3) (patchOl) {}; \node[align=left] at (7.25,-3) (patchOr) {};
    \filldraw[align=left, fill=gray!30, rounded corners=4] (5,-3.25) rectangle node (patchO)
        {\texttt{patch.o}} ++(2.25,0.5);

    \draw[align=left,draw=none](1,-4.5) rectangle node (dev) {\textbf{Virtual Chip}} ++(6,0.5); 
    \filldraw[darkblue, align=left, fill=darkblue!30, rounded corners=4](1,-4.5) rectangle node (asd) {} ++(6,-4.75); 
    \filldraw[align=left, fill=white](3,-5) rectangle node (buff1) {\texttt{patch.elf}} ++(2.0,0.5); 
    \draw [decorate,decoration={brace,amplitude=20pt}]
    ( 7,-4.5) -- ( 7,-9.25) node (patchELF) [black,midway,xshift=15] {};

    \filldraw[align=left, fill=white, dashed](3,-8.0) rectangle node (buff1) {} ++(2.0,0.5); 
    \filldraw[align=left, fill=white](3,-6  ) rectangle node (buff1) {\texttt{ROM}} ++(2.0,0.5); 
    \filldraw[align=left, fill=white](3,-6.5) rectangle node (buff1) {\texttt{RAM}} ++(2.0,0.5); 
    \filldraw[align=left, fill=white](3,-7.0) rectangle node (buff1) {\texttt{MMIO1}} ++(2.0,0.5); 
    \filldraw[align=left, fill=white](3,-7.5) rectangle node (buff1) {\texttt{MMIO2}} ++(2.0,0.5); 
    \filldraw[align=left, fill=white](3,-9.0) rectangle node (buff1) {\texttt{patch.c}\\\texttt{\_start()}} ++(2.0,1.0); 
    \node[align=right] at (1.75,-8.75) (ep) {Entry};
    \draw [->] (ep) -> node [above] {} (2.75,-8.75);

    \draw [decorate,decoration={brace,amplitude=10pt}]
    (3,-6.5) -- (3,-5.5) node (patchELFMod) [black,midway,xshift=-6] {};

    \node[align=right] at (5,-6.25) (cont) {};
    \node[align=left] at (5.5,-8.25) (patch2) {2.};
    \node[align=right] at (2.5,-8.25) (patch1) {1.};

    \draw [->] (firmware) -> node [above] {\texttt{xmit\_state}} (dumpBinl);
    \draw [->] (dumpBinr) -> node [above] {\texttt{objcopy}} (dumpOl);
    \draw [->] (dumpOb) -> (ldScriptst);
    \draw [->] (patchCr) -> node [above] {Compile} (patchOl);
    \draw [->] (patchOr) -> (ldScriptsl);

    \draw [->] (ldScriptsb)  |- node [above, pos=.75] {Link} (patchELF.east);
    \draw [->] (patch2)  -| node [rotate=90, anchor=west] {} ++(0.5,0) |- (cont.east);

    \node [rotate=270] at (6.25,-7.25) {\texttt{cont()}};
    \draw [->] (patch1)  -| node [sloped, above, anchor=center] {} ++(-0.5,0) |- (patchELFMod);
    \node [rotate=270] at (2.25,-7.25) {Modify};

\end{tikzpicture}
}
\caption{Reassembling the firmware image and live snapshot to an executable \ac{ELF} file.}
\vspace{-1em} 
\label{fig:patching}
\end{figure}

We use an unmodified \ac{QEMU} in user mode for emulation.
However, the snapshot is a raw binary without symbols.
We re-assemble it to an \ac{ELF} file, as illustrated in \autoref{fig:patching}.
User-defined code is then compiled and linked against the firmware image.
The compiled code is stored in a separate page and provides the initial entry point \texttt{\_start} for the emulation.
It shares the same address space as the firmware, hence it can call functions and parse data structures within the image.
The syntax is equivalent to any \emph{C} code written for the firmware. 
It also adds new features and makes modifications to substitute missing physical hardware.

\fuzztool runs in \emph{Linux} user mode, which does not support interrupts.
Thus, we disable functions responsible for enabling and disabling interrupts.
Timing-related functions, such as delay, use special purpose hardware and are also replaced.
\emph{ThreadX} uses a \ac{SVC} to perform a context switch between threads.
On \ac{ARM} this is a software interrupt, with a handler located at a known location.
As an \ac{SVC} has special calling conventions that cannot be emulated in user-mode, we re-implemented the handler.

After these modifications, the firmware is executed until the idle thread returns from the interrupt.\footnote{On \ac{ARM}, returning from exceptions is done by loading a special value to the \ac{PC}. The idle thread will return to \texttt{0xfffffffd}, showing that an interrupt invoked it.}
We replace that return address on the stack with a pointer to our own function.
Within this function, we can invoke interrupt handlers like a normal function call to preserve the threading behavior.
Thereby, we can inject \ac{HCI} traffic or Bluetooth frames, as described in \autoref{ssec:uart} and \autoref{ssec:inject}.

\subsection{Hooking for Portability}
\label{ssec:hooking}
We implement a lightweight hooking mechanism that can be used to modify the emulated firmware as well as the firmware running on the device.
Any code written in \fuzztool can also be compiled for the firmware and injected like a shellcode.
Even though the firmware is in ROM, it can be patched temporarily.
\emph{Broadcom} uses a Patchram mechanism to do so~\cite{mantz2019internalblue}. Each Patchram slot contains a \SI{4}{\byte} overlay in ROM and can be used to branch to the actual patch. The number of Patchram slots is very limited, but we use this mechanism as it allows us to install patches on the virtual and the physical firmware.

As the number of modifications to the ROM is limited to 256 Patchram slots on the \emph{CYW20735} chip, we use a trampoline-based approach,
similar to the \emph{Nexmon} hook patch variant~\cite{schulz2018}.
More advanced approaches like \emph{RetroWrite} that pose less overhead are completely infeasible, as they rewrite the whole firmware and require position-independent code~\cite{dinesh2020retrowrite}.
Instead, we modify the prologue of the target function to branch to our code.
Once our hook is executed, we restore the original prologue and call the target function.
On return, we execute a post-hook function to reinstall the hook and continue normal execution.

This hooking mechanism enables \fuzztool to trace function calls and analyze interrupt handlers and the corresponding status registers running on \ac{QEMU} and the physical device.
It also supports writing \acp{PoC} for over-the-air firmware vulnerabilities running on the physical hardware. 

For example, a basic \ac{LMP} protocol fuzzer requires the following hooks:

\begin{enumerate}
\setlength\itemsep{0.1em}
\item context switches between threads,
\item \acf{HCI} support,
\item hardware interrupt based timers, and
\item $\sim$100 hooks for debugging and implementation.
\end{enumerate}

\subsection{Heap Sanitizer Hook Performance}
\label{ssec:hsan}
\emph{ThreadX} has a custom implementation for dynamic memory called \texttt{BLOC} buffer.
Each \texttt{BLOC} is a continuous chunk of memory, divided into several chunks of equal size.
Free chunks are managed using a singly linked list.

The sanitizer iterates over the free list and validates that all pointers are  within the \texttt{BLOC} pool. \fuzztool hooks various functions such as \path{memcpy} and \path{dynamic_memory_Release} to integrate this check without further modifications to the heap itself. Thus, the \fuzztool sanitizer can also be added during runtime to the firmware running on the physical device.

\begin{figure}[!b]
\centering

\scalebox{0.7}{
    \begin{tikzpicture}
        \begin{axis}[        
        legend pos=north west,
        width=12cm,
        height=6cm,
	    boxplot/draw direction=y,
        ylabel={Runtime in seconds \emph{(log scale)}},
        ymode=log,
        ymin=0,
        ymax=0.3,
        ytick={0,0.025,0.05,0.1,0.2},
        yticklabels={0.025,0.05,0.1,0.2},
        xtick={1,2,3},
        xticklabel style={align=center},
        xticklabels={QEMU \\ baseline, \fuzztool with \\ heap sanitizer, \emph{Unicorn} with \\ heap sanitizer}
        ]
                \addplot+ [boxplot,color=blue!70] table [y index=1,col sep=comma] {measurements/plain.csv};
                \addplot+ [boxplot,color=red!70] table [y index=1,col sep=comma] {measurements/frankenstein_sanitizer.csv};
                \addplot+ [boxplot,color=gray] table [y index=1,col sep=comma] {measurements/uc_sanitizer.csv};

        \end{axis}
    \end{tikzpicture}
    }
\caption{Performance comparison of heap sanitizer with \fuzztool and \emph{Unicorn} hooks in LMP fuzzing.}
\label{fig:frankensteinperformance}
\end{figure}
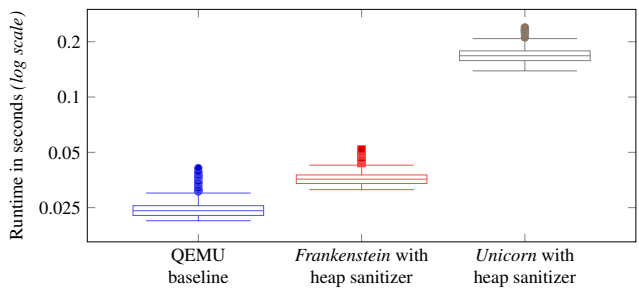

\emph{Unicorn}, which is the state-of-the-art firmware hooking tool, allows setting callbacks for each executed basic block, instruction, or memory access.
It relies on external function calls~\cite{unicorn}.
Since \fuzztool hooks are modifications to the firmware itself, no external libraries are called.
In addition, the \fuzztool hook payload is executed within the instrumented firmware and implemented in $C$.
Therefore, it outperforms the \emph{Unicorn} hooking mechanism.

We re-implement the same heap sanitizer with \emph{Unicorn} \emph{Python} bindings
for comparison and run it on a \emph{Thinkpad T430} with an \emph{i5-3320M} CPU.
\autoref{fig:frankensteinperformance} shows the results. 
The baseline runtime of the instrumented firmware without heap sanitizer is \SI{24.8}{\milli\second} on average. \todo{mean or median?}
When sanitizing the heap during \ac{LMP} fuzzing, \fuzztool comes with a performance overhead of \SI{11.6}{\milli\second} (\SI{46.8}{\percent}) on the mean average compared to the baseline.
The same implementation using \emph{Unicorn} increases the runtime by \SI{145.2}{\milli\second} (\SI{585.5}{\percent}) compared to the baseline. 
Therefore, firmware instrumentation using \fuzztool outperforms \emph{Unicorn} by a factor of $12.5$ in the heap sanitizer scenario. 
Performance of other use cases varies depending on the number of hooks.

While the exact speedup depends on the scenario, it is sufficient to overcome the break-even point for the
full-stack fuzzing use case. 
\fuzztool emulates the firmware fast enough to enable interaction with an unmodified Bluetooth stack on the host and, thus, attaching it to \emph{Linux BlueZ}~\cite{bluez}.

\subsection{Talking to an Operating System}
\label{ssec:uart}
Attaching \fuzztool to an operating system Bluetooth implementation enables full-stack fuzzing. For example, \cvea (see \autoref{ssec:cvea}) is triggered by the host asking for additional information. 
On the physical device, \ac{HCI} traffic is sent to the host via \ac{UART}.
In the emulation, we connect \ac{UART} to a \emph{Linux} host using a pseudo-terminal device~\cite{ptmx}.
Opening a \ac{PTM} creates a file descriptor, used in the emulator via \emph{Linux} \path{read} and \path{write} system calls.
The operating system then creates a corresponding \ac{PTS}, which is similar to a virtual serial interface. 
The \ac{PTS} is then passed to \texttt{btattach} to attach the emulator to the \emph{Linux BlueZ} Bluetooth stack.

\ac{HCI} events generated by the firmware are extracted by hooking \path{uart_SendAsynch}.
This function is a central component of the transmit state machine and gets called for every \ac{HCI} event. 
Those events are sent to the host using the \path{write} system call.
The opposite direction, injecting \ac{HCI} commands, requires two steps.
We replace functions that read data from \ac{UART} packet buffers with \path{read} system calls and analyze the status registers triggering the \ac{UART} interrupt handler.
This will invoke the \ac{UART} receive state machine implemented in the \path{bttransport} thread.
Note that \ac{ACL} and \ac{SCO} data traffic is also passed over the \ac{UART} interface.

\subsection{Non-Wireless Wireless Packet Injection}
\label{ssec:inject}

The virtual modem calls the \acf{BCS} interrupt handler and generates specific packets for these.
For most task types, the packets can be entirely random for reaching maximum coverage.
The \acf{LMP} was fuzzed coverage-based due to the complexity of the \acf{LM} state machine.
The most interesting fuzzing optimizations are as follows.

\begin{description}
\item[Paging] This task accepts any connection attempt.
\item[LMP] The \ac{LM} handles a lot of logic within the firmware and is fuzzed coverage-based, as described in~\autoref{ssec:coverage}.
\end{description}

\begin{table}[!b]
\caption{Calling convention for an \acs{ACL} slave connection.}
	\scriptsize
    \label{tab:phystat}
    \centering
    \renewcommand{\arraystretch}{1.3}
    \begin{tabular}{|l|l|}
    \hline
    \textbf{Bluetooth Clock} & \textbf{\path{phy_status}}\\
    \hline
    \texttt{0b\textcolor{gray}{??}00} & Receive header done \\
    \texttt{0b\textcolor{gray}{??}01} & Receive done, \emph{Slot01} interrupt \\
    \texttt{0b\textcolor{gray}{??}10} & Transmit done \\
    \texttt{0b\textcolor{gray}{??}11} & \emph{Slot11} interrupt \\

	\hline
\end{tabular}
\end{table}

We have to analyze the calling convention of \texttt{bluetooth} 
\texttt{CoreInt\_C} to implement a virtual modem injecting custom packets.
On the device, it is important not to alter the \SI{312.5}{\micro\second} timing at which \path{bluetoothCoreInt_C} is called.
Hence, only a limited number of debugging techniques can be used.
We hook this function on the physical device and dump the hardware registers of interest to a ring buffer.
Those are mainly \path{phy_status} and \path{sr_status}, which control the \ac{BCS} kernel.
\path{phy_status} controls which function is executed by \path{bluetoothCoreInt_C} and depends on the Bluetooth clock.

An example of \path{phy_status} within an active \ac{ACL} slave connection is shown in \autoref{tab:phystat}.
Prior to a reception in the \emph{Slot11} interrupt, the receive buffer located in RAM is mapped to the hardware receive buffer using \ac{DMA}.
This memory overlay technique is used to prevent the use of \path{memcpy} and therefore save CPU resources.
Packet data is written to RAM instead of writing it directly to the hardware receive buffer. 
Within the \emph{receive header done} interrupt, the packet header is available.
Besides, it is checked whether the remote device acknowledged the previous transmission.
If no retransmission is required, the next packet is put into the \ac{ACL} task storage for transmission.
Those \ac{LMP} packets which the remote device acknowledged are passed back to the \ac{LM} for final processing. 
Once the packet has been received, \emph{receive done} is called.
The receive buffer is unmapped, and the packet is saved to the \ac{ACL} task storage. 
During the \emph{Slot01} interrupt, the hardware is configured to transmit the next packet.
In addition, the received packet is handed to the corresponding thread.
\ac{LMP} packets are passed to the \ac{LM} thread. \ac{ACL} packets are passed to the \texttt{bttransport} thread.
\emph{Transmit done} will unmap the transmit buffer.
This process repeats with the \emph{Slot11} interrupt.

\begin{figure}[!b]
\vspace{-1em} 
\centering
\scalebox{0.7}{
  \begin{tikzpicture}
    \begin{axis}[
        xlabel={Executed test cases \emph{(log scale)}},
        ylabel={Coverage in basic blocks},
        ymin=1000,
        ymax=4000,
        xmin=1,
        xmax=2000000,
        legend pos=north west,
        width=12cm,
        height=7cm,
        xmode=log,
      ]
      \addplot[color=blue!70,thick] table [col sep=comma] {measurements/coverage_protofuzz_HCI.csv};
      \addplot[color=red!70,densely dashed] table [col sep=comma] {measurements/coverage_protofuzz.csv};
      \addplot[color=gray,densely dotted] table [col sep=comma] {measurements/coverage_randomFuzz.csv};
      \legend{Packet level with \emph{BlueZ}, Packet level, Classic BLOB}

    \end{axis}
  \end{tikzpicture}
}
\vspace{-1.5em}
\caption{LMP fuzzing strategy comparison.}
\label{fig:cov_fuzz_comp}
\end{figure}
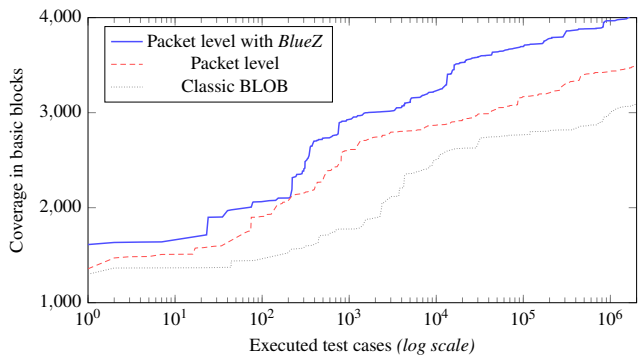

\begin{table}[!b]
\caption{Coverage increase by new zero-click attack surfaces.}
\vspace{-0.5em}
    \label{tbl:cov_inc}
    \centering
    \scriptsize
    \renewcommand{\arraystretch}{1.3}
    \begin{tabular}{|l|l|}
    \hline
    \textbf{Scenario} & \textbf{Coverage} \\
    \hline
    LMP fuzzing &  \SI{2.76}{\percent} \\
    LMP fuzzing with \emph{BlueZ} \ac{HCI} &  \SI{0.56}{\percent} \\
    Attach to stack  \texttt{hciconfig hci1 up} & \SI{2.04}{\percent} \\
    Attach to stack  \texttt{hcitool scan} & \SI{0.59}{\percent} \\
    Attach to stack  \texttt{hcitool cc} & \SI{1.04}{\percent} \\
    Attach to stack  \texttt{hcitool lescan} & \SI{0.57}{\percent} \\
    Attach to stack  \texttt{hcitool lecc} & \SI{1.85}{\percent} \\
    \hline
    \textbf{Total} & \textbf{\SI{9.40}{\percent}} \\
    \hline
    \end{tabular}
\end{table}

\subsection{Code Coverage}
\label{ssec:coverage}

We came up with a different representation for coverage-guided fuzzing of protocols.
Instead of handling all the input data as a single \ac{BLOB}, we represent it as a sequence of packets, where packets and sequences are mutated separately.
This enables the fuzzer to reorder already known packets to increase coverage.
We start with a single sequence containing only one packet that consists of null-bytes.
For each execution, a random sequence is chosen from the population and mutated.
To distinguish the effect of mutating sequences from mutating packets, only one of those is performed per measurement.
If the mutation of a single packet increased the code coverage, the sequence containing the new packet is added to the population.
If mutating a sequence increased the code coverage, it is also added to the population.
Sequence mutations are reordering packets, inserting known packets from other sequences, or merging two sequences.

Both approaches were compared with the same set of mutations over two million test cases, as shown in \autoref{fig:cov_fuzz_comp}.
We previously implemented this reference implementation~\cite{randomFuzz}.
Our adaptive approach finds more blocks in a shorter amount of time.
The total coverage for \ac{LMP} fuzzing converges to \SI{2.76}{\percent}.
Introducing \ac{HCI} support increases the coverage further by \SI{0.56}{\percent}, as \ac{HCI} handlers and the \ac{UART} receive state machine are invoked.

We evaluate the total code coverage during \ac{LMP} and \ac{BCS} task fuzzing.
This was obtained by using \ac{QEMU} with the \path{translate_block} trace option.
The total code coverage is then loaded to \emph{IDA Lighthouse} plugin~\cite{lighthouse} to determine the percentage coverage shown in \autoref{tbl:cov_inc}.
Each row shows the amount of new code reached using the described method.

The total code coverage we reached so far is \SI{9.40}{\percent}.
However, we only analyzed specific scenarios prior to pairing, which enable potential zero-click attacks.
This focus is reasonable as the Patchram is limited and \emph{Broadcom} will likely not fix issues that require pairing.
The code coverage reached is comparable to the size of the related parts within the Bluetooth specification.
For example, we reached \SI{3.32}{\percent} code coverage by fuzzing \ac{LMP}, and the chapter containing \ac{LMP} in the specification is only \SI{4.05}{\percent} of the total Bluetooth specification in pages~\cite[p. 567ff]{bt52}. Also, \emph{Broadcom} provides vendor-specific additions and utilizes the \emph{ThreadX} operating system, which are not part of the specification.

As also shown in \autoref{tbl:cov_inc}, coverage increases by attaching the firmware to the host
stack. This realistic behavior is possible due to \fuzztool's full-stack approach.
Fuzzers that do not implement thread switches and only focus on one specific protocol handler
cannot reach these protocol parts by design.
This includes \cvea that requires interaction between the \emph{BCS} kernel and link manager.
Moreover, as \fuzztool includes host stack behavior, the identified issues will
likely reproduce on physical devices.

Coverage also offers further insights. Even with the partial symbols, identifying relevant functions is complex.
Simply calling a function in emulation and observing the execution
can help to gain valuable high-level insights into the code.
For example, \num{420} functions end on \path{Rx} and potentially receive
data. Observing coverage enables us to determine which of these functions are important
and in which reception handler context they are called.

\subsection{Adding New Firmware}
Apart from this use case described here, \fuzztool is also capable of fully emulating firmware if no memory snapshot
is available but only the compiled firmware including debug symbols.
\ac{ELF} is a common format of these images that can be directly imported into \fuzztool.
Without a memory snapshot, hardware initialization needs to be performed, which is challenging
in complex environments. In the use case not described here we were able to set up a working
emulation within half a week, including support of buttons, \ac{SPI}, and \ac{CAN} interfaces of 
a smaller firmware.

Our workflow for integrating new firmware looks as follows.
The firmware is executed until a fault---such as infinite loop or illegal instruction--occurs.
Then, we fix the root cause of this.
We add function tracing hooks to function calls that seem to be relevant.
Those function calls are displayed during emulation to show the program flow.
Prior to functions or interrupt handlers, hardware registers and buffers can be modified, e.g., using \path{read}.
This includes clock values and receive buffers of external hardware.
Then, coverage-guided fuzzing can be used to verify how the input is processed by the firmware.


\section{Fuzzing Results and Exploitation}
\label{sec:sec_analysis}
This section describes the heap exploitation technique and documents three heap overflows.

\subsection{Heap Corruption}
None of the observed devices implements any exploit mitigation, such as \ac{DEP} or \ac{ASLR}.
The memory allocator described in \autoref{ssec:hsan} can be easily exploited.
With a heap overflow, an attacker can control a free list pointer to point to any location.
This pointer is treated as a valid \texttt{BLOC} buffer due to repetitive allocations, as depicted in \autoref{fig:free_ovf_free}.
This leads to a \emph{write-what-where} gadget and allows for \acf{RCE}.
The technique has already been discussed for the exploitation of \emph{Marvell} Wi-Fi controllers, although it was not used in the actual exploit~\cite{marvell2019}.

\subsection{Classic Bluetooth Device Scanning EIR (\cvea)}
\label{ssec:cvea}
This section describes a heap overflow exploit in device inquiry, utilizing the full stack~\cite[p. 513]{bt52}. 
As a device scans for other devices, these can respond with an \ac{EIR}.
An \ac{EIR} contains additional information such as the device name, which is copied into an \ac{HCI} event to be displayed to the user to list available devices for pairing. The \ac{EIR} length is extracted from the payload header and subject to the same physical-layer constraints such as data rate and maximum packet duration. Due to these physical-layer restrictions, the firmware skips further length checks prior to copying an \ac{EIR}.

\tikzset{>=latex}

\begin{figure}[!b]
    \centering
    \subfloat[Layout immediately after a heap overflow.]{\scalebox{0.7}{\BlockOverflowImageOvf}
    }

    \centering
    \subfloat[Layout after a heap overflow and a free of the affected buffer.]{
    \scalebox{0.7}{\BlockOverflowImageFree}
    }

    \caption{Effect of overflowing a free \texttt{BLOC} buffer.}
    \label{fig:free_ovf_free}
\end{figure}

\begin{figure}[!b]
\centering
\scalebox{0.7}{
\begin{tikzpicture}

    \node at (6,1.5) {Validated in hardware};
    \node at (10.5,1.5) {Should be zero};
    
    \draw [decorate,decoration={brace,amplitude=10pt}]
    (3.25,0.7) -- (8.75,0.7) node (lenb) [black,midway,xshift=-6] {};
    \draw [decorate,decoration={brace,amplitude=10pt}]
    (9.25,0.7) -- (11.75,0.7) node (lenb) [black,midway,xshift=-6] {};
    
    \filldraw[align=left, fill=white](0,0) rectangle node (llid) {LLID} ++(2.0,0.5); 
    \draw[align=center](1,-0.3) node (llidt) {\textcolor{gray}{\SI{2}{\byte}}}; 
    \filldraw[align=left, fill=white](2,0) rectangle node (flow) {Flow} ++(1.0,0.5);
    \draw[align=center](2.5,-0.3) node (flowt) {\textcolor{gray}{\SI{1}{\byte}}}; 
    \filldraw[align=left, fill=white](3,0) rectangle node (len) {Length} ++(6.0,0.5);
    \draw[align=center](6,-0.3) node (lent) {\textcolor{gray}{\SI{10}{\byte}}}; 
    \filldraw[align=left, fill=red!30](9,0) rectangle node (rfu) {RFU} ++(3.0,0.5);
    \draw[align=center](10.5,-0.3) node (ruft) {\textcolor{gray}{\SI{3}{\byte}}}; 
    
\end{tikzpicture}
}
\vspace{-1em} 
\caption{Payload header format for multi-slot \ac{ACL} packets and all \ac{EDR} \ac{ACL} packets.}
\label{fig:multislotheader}
\end{figure}

\begin{figure}[!b]
\begin{lstlisting}[linewidth=0.97\columnwidth]
@\hspace{-0.5em}\vspace{-0.25em}\colorbox{white}{25 25 25 25 25 25 25 25 25 25 25 25}@ @\hspace{-0.95em}\colorbox{gray!30}{ca fe ba be}@ @\textcolor{gray!80}{\textrm{MAC address}}@
@\hspace{-0.5em}\vspace{-0.25em}\colorbox{gray!30}{be ef}\hspace{-0.3em}@ @\hspace{-0.64em}\colorbox{blue!30}{e0 04 c8 01 02 03 04 05 06 07 08 09 0a 0b}@ @\textcolor{blue!80}{\textrm{EIR}}@
@\hspace{-0.5em}\vspace{-0.25em}\colorbox{blue!30}{0c 0d 0e 0f 10 11 12 13 14 15 16 17 18 19 1a 1b}@ 
@\hspace{-0.5em}\vspace{-0.25em}\colorbox{blue!30}{.. .. .. .. .. .. .. .. .. .. .. .. .. .. .. ..}@ 
@\hspace{-0.5em}\vspace{-0.25em}\colorbox{blue!30}{cc cd ce cf d0 d1 d2 d3 d4 d5 d6 d7 d8 d9 da db}@ 
@\hspace{-0.5em}\vspace{-0.25em}\colorbox{blue!30}{dc dd de df}\hspace{-0.35em}@@\colorbox{white}{f4 d6 41 9a}@@\hspace{-0.35em}\colorbox{red!30}{64 65 66 67 68 69 6a 6b}@ @\textcolor{red!80}{\textrm{Duplicated}}@
@\hspace{-0.5em}\vspace{-0.25em}\colorbox{red!30}{6c 6d 6e 6f 70 71 72 73 74 75 76 77 78 79 7a 7b}@ @\textcolor{red!80}{\textrm{suffix}}@
@\hspace{-0.5em}\vspace{-0.25em}\colorbox{red!30}{7c 7d 7e 7f }@@\hspace{-0.75em}\vspace{-0.25em}\colorbox{white}{00 26 26 26 26 26 26 26 26 26 26 26}@
\end{lstlisting}
\vspace{-1em} 
\caption{Overflowing hardware buffer that allows control over more bytes than the actual payload length.}
\label{fig:hwbuf}
\end{figure}

\autoref{fig:multislotheader} shows the \ac{ACL} header format~\cite[p.~482]{bt52}.
The packet length is followed by \ac{RFU} bits, which should be set to zero. The firmware  includes these bits in the packet length.
Non-zero \ac{RFU} bits exceed the buffer length of the \ac{HCI} event. 

The hardware buffer holding the payload is not restricted to the payload length of the specific packet being parsed. Even worse, it contains a duplicate of the packet payload, as depicted in \autoref{fig:hwbuf}. This makes memory located after the original packet's payload predictable.

We allocate three buffers in a row within the affected \texttt{BLOC} pool to
 exploit this heap overflow with a write-what-where gadget.
This cannot be achieved using the \ac{EIR} packets, as the data rate is too low compared to the \ac{UART} connection to the host---the \texttt{BLOC} pool would be cleared faster than filled.

We exploit that the host issues an \path{HCI_Remote_Name_Request} command when an unknown device connects~\cite[p. 1815ff]{bt52}.
The returned \path{HCI_Remote_Name_Request_Complete} event has the correct size to be allocated in the affected \texttt{BLOC} pool.
The attacker-controlled remote name is read via \ac{LMP} in multiple packets into that buffer.
By omitting the last packet and silently dropping the connection, the buffer is kept for several seconds until a timeout occurs.
Repeating this process, we can write arbitrary memory, resulting in \ac{RCE}.

The over-the-air \ac{PoC} works on various devices, as listed in \autoref{tab:eirvuln}.
By overflowing the \texttt{BLOC} header with an invalid address, the Bluetooth chip of the device under test crashes.
The \ac{PoC} running on the \emph{CYW20735} evaluation board changes the device name to the payload and \ac{MAC} address to pretend to be multiple physical devices.
This method works well against \emph{Android} and \emph{Linux} hosts.

\begin{table}[!t]
\caption{Devices vulnerable to \cvea.}
    \label{tab:eirvuln}
    \centering
    \scriptsize
    \renewcommand{\arraystretch}{1.3}
    \begin{tabular}{|l|p{3.3cm}|p{1.4cm}|l|l|}
    \hline
    \textbf{Chip} & \textbf{Device}               & \textbf{Build Date}  & \textbf{Vuln}\\
    \hline
    BCM20702           & Thinkpad T430                      &  $<$ 2010? &Yes\\
    BCM4335C0          & Nexus 5, Xperia Z3 Compact, \newline Samsung Galaxy Note 3, LG G4    &  Dec 11 2012 &Yes\\
    BCM4345B0		   & iPhone 6 (unfixed in \emph{iOS 12.4}) & Jul 15 2013 & Yes \\
    BCM4358A3          & Samsung Galaxy S6, Nexus 6P        &  Oct 23 2014 &Yes\\
    BCM4345C1		   & iPhone SE (prior \emph{iOS 12.4})  & Jan 27 2015 & Yes \\
    Unknown            & Samsung Galaxy A3 (2016)           & Unknown     &Yes\\
    BCM20707           & Fitbit Ionic                       & Unknown     &Yes\\
    BCM4347B0          & Samsung Galaxy S8                  & Jun 3 2016  &Yes\\
    BCM4347B1 & iPhone 8/X/XR (prior \emph{iOS 12.4}) & Oct 11 2016 & Yes \\
    BCM4357            & Samsung Galaxy 9+/Note 9           & Unknown     &Yes\\
    CYW20735B1         & Evaluation Board                   & Jan 18 2018 &Yes\\
    BCM4375B1          & Samsung Galaxy S10e/S10/S10+       & Apr 13 2018 &No \\
    CYW20819A1         & Evaluation Board                   &  May 22 2018 &Yes\\
	\hline
\end{tabular}
\end{table}

The Bluetooth stack on \emph{Apple} devices does not allow for multiple unauthenticated connections simultaneously and is not covered by our \ac{PoC}.
We extracted ROM and Patchram from jailbroken \emph{iOS 12.4} devices with \emph{InternalBlue} and can confirm that the \emph{iPhone SE}, \emph{7}, and \emph{8/X/XR} received a patch in August 2019 or earlier. On the \emph{iPhone 6}, the vulnerability is still unpatched, but all Patchram slots are already occupied.

Since \emph{Android} needs to support a lot of different hardware, and vendors need to apply individual fixes, patches take a bit longer. A fix was issued on August 5 2019, and it took \emph{Samsung} until mid-September to roll out these patches for their devices.

The \ac{EIR} vulnerability requires users to scan for devices.
We were able to observe device scanning in practice, for example, every few hours inside a residential accommodation.
However, we do not know which apps or user actions did trigger device scanning.
Some apps require frequent scanning.
Bhaskar et al. built a smartphone app used by law enforcement that scans for credit card skimmers using classic Bluetooth~\cite{bhaskar2019}.
In our observations, location services only use \ac{BLE} device scanning, but no classic device scanning.

\subsection{Any BLE Packet (\cveb)}

\begin{figure}[!t]
\centering
\scalebox{0.7}{
\begin{tikzpicture}

    \path[draw=gray] (11.75,-0.5) -- (11.75,-1.1);
    \path[draw=gray] (10.75,1.1) -- (10.75,-1.1);
    \path[draw=gray] (10.25,-0.5) -- (10.25,-1.1);
    \path[draw=gray] (8.75,1.1) -- (8.75,-1.1);
    \path[draw=gray] (3,-0.5) -- (3,-1.1);
    \path[draw=gray] (1.75,1.1) -- (1.75,-1.1);
    \path[draw=gray,dashed] (0.3,0) -- (2,0);
    
    \draw[align=left, draw=none](0,0) rectangle node (bloc) {\texttt{BLOC}} ++(2.0,0.5);
    \filldraw[align=left, fill=blue!30](1.75,0) rectangle node (buf) {Buffer} ++(7,0.5);
    \filldraw[align=left, fill=red!30](8.75,0) rectangle node (buf) {Pointer} ++(2,0.5);
    \draw[align=center](5.5,0.8) node (bufft) {\textcolor{gray}{\SI{264}{\byte}}}; 
    \draw[align=center](9.75,0.8) node (headt) {\textcolor{gray}{\SI{4}{\byte}}}; 
    
    \draw[align=left, draw=none](0,-0.5) rectangle node (packet) {Packet} ++(2.0,0.5);
    \filldraw[align=left, fill=white](3,-0.5) rectangle node (pdu) {Packet Data Unit (PDU)} ++(7.25,0.5);
    \filldraw[align=left, fill=white](10.25,-0.5) rectangle node (pdu) {CRC} ++(1.5,0.5);
    \draw[align=center](2.325,-0.8) node (empty) {\textcolor{gray}{\SI{12}{\byte}}}; 
    \draw[align=center](6,-0.8) node (packett) {\textcolor{gray}{\SI{252}{\byte}}}; 
    \draw[align=center](9.5,-0.8) node (pdut2) {\textcolor{gray}{\SI{3}{\byte}}}; 
    \draw[align=center](10.5,-0.8) node (crct1) {\textcolor{gray}{1}}; 
    \draw[align=center](11.3,-0.8) node (crct2) {\textcolor{gray}{\SI{2}{\byte}}}; 
    
\end{tikzpicture}

}
\vspace{-0.5em}
\caption{\ac{BLE} PDU violating a \texttt{BLOC} buffer.}
\vspace{-1em}
\label{fig:blepdu}
\end{figure}
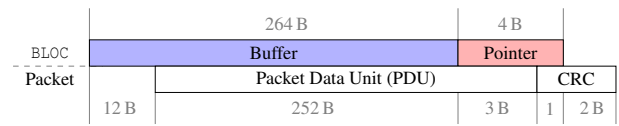

This section describes a heap overflow in the reception of \ac{LE} \acp{PDU}. In Bluetooth 4.2 the maximum \ac{PDU} length was extended from \SI{20}{\byte} to \SI{255}{\byte}. 
A \ac{PDU} is stored in a special purpose \texttt{BLOC} pool with a buffer size of \SI{264}{\byte}, as shown in \autoref{fig:blepdu}.
Besides the \ac{PDU} payload, the buffer also contains  \SI{12}{\byte} for headers.
Therefore, the buffer is \SI{3}{\byte} too small to hold the maximum total \ac{PDU} length of \SI{255}{\byte}.
These \SI{3}{\byte} are part of the pointer to the next free \texttt{BLOC} buffer, as previously depicted in \autoref{fig:free_ovf_free}.

However, the fourth overflowing byte is determined by the \ac{CRC} and also copied.
It is stored in the receive buffer, despite previously being validated in hardware.
An attacker has to adapt the payload, including the \ac{CRC}, to take control over the heap.
The initial \ac{CRC} state is randomized for each connection.
Malicious packets with a chosen \ac{CRC} need to be produced within the tight Bluetooth clock to prevent connection termination.
The attacker can pre-calculate the header and first \SI{248}{\byte} payload. The payload can be static for this attack. The next \SI{4}{\byte} payload are used to adjust the \ac{CRC}.
After this, \SI{3}{\byte} of the \texttt{BLOC} buffer header are inserted. These are followed by the chosen \ac{CRC}, which manipulates the remaining \SI{1}{\byte} of the header.

\begin{table}[!t]
\caption{Devices vulnerable to \cveb.}
    \label{tab:crcvuln}
    \scriptsize
    \centering
    \renewcommand{\arraystretch}{1.3}
    \begin{tabular}{|l|p{3.3cm}|p{1.4cm}|l|l|}
    \hline
    \textbf{Chip} & \textbf{Device}               & \textbf{Build Date}  & \textbf{Vuln}\\
    \hline
    $<$ Bluetooth 4.2    & ---               & $<$ 2014 & No \\
    BCM4345C0          & Raspberry Pi 3+/4        &  Aug 19 2014 &Yes\\
    BCM4347B0          & Samsung Galaxy S8                  & Jun 3 2016  &Yes\\
    CYW20719B1         & Evaluation Board         & Jan 17 2017 &Yes\\
    CYW20735B1         & Evaluation Board         & Jan 18 2018 &Yes\\
    BCM4375B1          & Samsung Galaxy S10e/S10/S10+          & Apr 13 2018 &Crash \\
    CYW20819A1         & Evaluation Board         &  May 22 2018 &Yes\\
	\hline
\end{tabular}
\end{table}

\begin{figure*}[h]
\center
\includegraphics[width=0.63\textwidth]{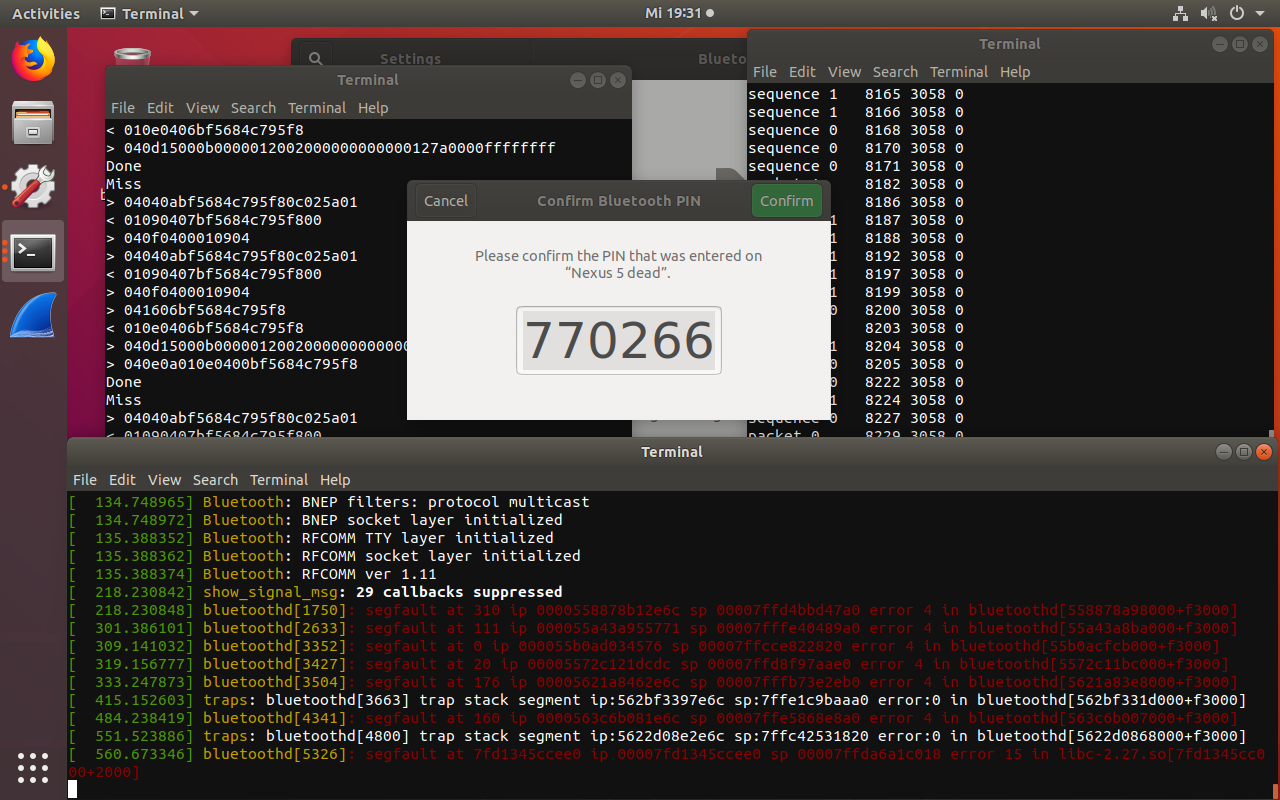}
\caption{LMP fuzzing results on \emph{Ubuntu} with \emph{BlueZ}.}
\label{fig:lmpubuntu}
\end{figure*}

Our current \ac{PoC} works over-the-air, but only allocates two \texttt{BLOC} buffers at once by sending a fragmented \ac{GATT} notification within one \ac{BLE} event. One additional buffer allocation is needed to gain \ac{RCE} with a write-what-where gadget.
Since the affected \texttt{BLOC} buffer is one of the largest, we assume that there is a standard-compliant way to execute this attack, i.e., using the 1M or 2M PHY modes.
\autoref{tab:crcvuln} shows a list of tested devices, which was validated with the partial \ac{PoC} and local buffer debugging on the device using \emph{InternalBlue}.

Interestingly, the \emph{Samsung Galaxy S10e} is differently affected by exactly the same heap corruption.
Bluetooth crashes over-the-air with our \ac{PoC} because a new heap check was introduced.
It checks for overflows by saving the \ac{LR} and a static \SI{1}{\byte} canary at the end of each \texttt{BLOC} buffer element. If the check fails, it crashes gracefully. 
When this happens, only one heap element is allocated, and we could not deploy a write-what-where gadget. The heap check protects from \ac{RCE} with \cveb despite the bug still being present.
We were able to produce correct data for the heap check, which already requires all \SI{4}{\byte} of our overflow.
To control the next element header, an \SI{8}{\byte} overflow would be required.
Such a new \ac{RCE} might be found by either patching \cveb manually on the \emph{CYW20735} firmware and continue fuzzing with \fuzztool or by porting it to the non-vulnerable \emph{Samsung Galaxy S10e} firmware.

As we did not provide a full \ac{PoC} and \emph{Patchram} is limited, \cveb has not been fixed on any \ac{RCE} exploitable device to the best of our knowledge, despite reporting it in July 2019.

\subsection{Any ACL Packet (\cved)}
\label{ssec:evalbuffer}

Within classic Bluetooth, \acf{ACL} mode is used for data transfer, such as tethering or music streaming. Similar to \ac{HCI}, it is sent to the host using \ac{UART}, but with a different data prefix.

Upon driver initialization by the operating system, the Bluetooth chip signals the maximum packet and buffer size using the \path{HCI_Read_Buffer_Size} command~\cite[p. 795]{bt52}. Broadcom chips are configured for an \ac{ACL} length of \SI{1021}{\byte} and \num{8} packets. If this buffer is exceeded, this causes a heap overflow. It is important to note that this overflow cannot be exploited without bypassing the driver and operating system Bluetooth stack, which requires privileged access either way.

Yet, on the \emph{CYW20735} chip only, there is a buffer misconfiguration that makes \ac{ACL} exploitable.
The global variables \path{BT_ACL_HOST_TO_DEVICE_DEFAULT_SIZE} and \path{BT_ACL_DEVICE_TO_HOST_DEFAULT_SIZE} are set to \SI{384}{\byte}, while
the chip still signals a size of \SI{1021}{\byte} to the host. Thus, just setting up a regular headset for audio streaming as a user immediately results in a heap overflow. As the misconfiguration affects both directions, the heap overflow can also be triggered over-the-air by sending a few \emph{L2Ping} packets exceeding \SI{384}{\byte}. When reconfiguring the buffer size in \emph{WICED Studio 6.2}, this bricks the board's capability of flashing new firmware.

This vulnerability stopped us from further \ac{ACL} fuzzing with the emulated \emph{CYW20735} firmware.
It is impossible to take a snapshot during music streaming or tethering before the firmware crashes.
However, the \emph{CYW20819} firmware does not have this issue---and \fuzztool is almost completely ported to this newer firmware as of June 2020.

\subsection{BlueFrag (\emph{CVE-2020-0022})}

Nonetheless, we tried to create a \ac{PoC} for \cved based on the assumption that a chip might cache
\ac{ACL} packets if sent using \ac{L2CAP} fragments. Instead of crashing the chip, it crashed
within \path{bluetoothd} of an up-to-date \emph{Samsung Galaxy S10e} as of November 2019.
After the report, which contained a \ac{PoC} including a \ac{CFI} bypass to create a reverse shell
using Bluetooth within \SI{2}{\minute}, this was fixed in the \emph{Android} February 2020 patches as \emph{CVE-2020-0022}.
The details of this are covered in our blog post~\cite{bluefrag}.

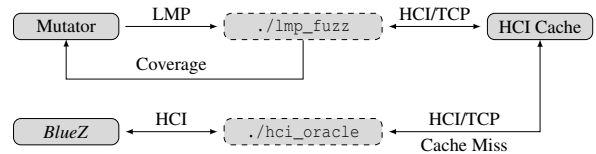
\begin{figure}[!t]
\centering
\scalebox{0.7}{
\begin{tikzpicture}
    \node[align=left] at (2,-0.25) (mutr) {};
    \filldraw[align=left, fill=gray!30, rounded corners=4](0,0) rectangle node (mut) {Mutator} ++(2,-0.5); 
    \node[align=left] at (4,-0.25) (fuzzl) {}; \node[align=left] at (7,-0.25) (fuzzr) {};
    \filldraw[align=left, fill=gray!30, dashed, rounded corners=4](4,0) rectangle node (fuzz) {\texttt{./lmp\_fuzz}} ++(3,-0.5); 
    \node[align=left] at (9,-0.25) (cachel) {}; 
    \filldraw[align=left, fill=gray!30, rounded corners=4](9,0) rectangle node (cache) {HCI Cache} ++(2,-0.5); 

    \node[align=left] at (4,-2.25) (oraclel) {}; \node[align=left] at (7,-2.25) (oracler) {}; 
    \filldraw[align=left, fill=gray!30, dashed, rounded corners=4](4,-2) rectangle node (oracle) {\texttt{./hci\_oracle}} ++(3,-0.5); 
    \node[align=left] at (2,-2.25) (bluezr) {};
    \filldraw[align=left, fill=gray!30, rounded corners=4](0,-2) rectangle node (bluez) {\emph{BlueZ}} ++(2,-0.5); 

    \node at (3,-1.0) {Coverage};
    \draw [->] (fuzz) |- node [sloped, above, anchor=center] {} ++(0,-1) -| (mut);
    \draw [->] (mutr) -> node [above] {LMP} (fuzzl);
    \draw [<->] (fuzzr) -> node [above] {HCI/TCP} (cachel);
    \draw [<->] (cache) |- node [above, pos=0.75] {HCI/TCP}  node [below, pos=0.75] {Cache Miss} (oracler);
    \draw [<->] (oraclel) -> node [above] {HCI} (bluezr);
\end{tikzpicture}
}
\caption{Testing of emulated LMP fuzzing against the \emph{Linux} \emph{BlueZ} Bluetooth stack.}
\label{fig:lmpfuzz}
\vspace{-1em} 
\end{figure}

\subsection{Link Management Protocol State Failures}

The \acf{LMP} in classic Bluetooth is managing connection and encryption setup. The protocol itself is rather simple.
However, the most recent attacks affecting a large fraction of Bluetooth devices were located in the \ac{LM} logic~\cite{knob, 2018:biham}.
Each packet type has a fixed length, with the maximum length being \SI{17}{\byte}~\cite[p. 679]{bt52}.

We attach the emulated firmware to a \emph{Linux} host to systematically test \ac{LMP} with \fuzztool, as depicted in \autoref{fig:lmpfuzz}.
The firmware processes \ac{LMP} packets generated by coverage-guided fuzzing, which in turn causes valid \ac{HCI} events.
A cache answers known event sequences, and unknown sequences are forwarded to the \emph{Linux} \emph{BlueZ} host implementation.
This differs from code coverage based tools like \texttt{syzkaller}~\cite{syzkaller}, because only valid management-related events are passed to the host. Moreover, we aim at increasing coverage within the firmware and not within the host.

This interplay with a real system generates various interesting outputs, as depicted in \autoref{fig:lmpubuntu}. The user interface shows a lot of weird pairing requests. We even observed faults that produced \texttt{dmesg} error outputs and one system freeze. However, they were hard to debug in practice, and we were not able to file specific bug reports.


\section{Discussion}
\label{sec:discussion}

This section discusses \fuzztool and patching of discovered vulnerabilities on a broader scope.
\autoref{ssec:applicability} provides an overview of other firmwares that could be fuzzed with \fuzztool. We show the current state of \emph{Broadcom} firmware patches on multiple generations of devices in \autoref{ssec:patching}. Mitigation techniques against our attacks are discussed in \autoref{ssec:memprotect} and \autoref{ssec:threadxheap}.

\subsection{Applicability to Other Systems}
\label{ssec:applicability}

The general idea of emulating firmware to facilitate wireless fuzzing can also be applied to other chips. An emulator similar to \ac{QEMU} and a basic understanding of the firmware binary are required, though. 

Our emulation framework is tailored to \ac{ARM} chips and \emph{ThreadX}.
\emph{ThreadX} is the number one \ac{RTOS}, which runs on over \SI{6.2}{\billion} devices and provides multiple ARM implementations~\cite{threadx}. Wireless firmware designed for this combination is wide-spread.
The other firmware that we internally ported for \fuzztool is \ac{ARM}-based and does not use
any operating system at all.

In the following, we provide an overview of wireless firmware based on similar technologies.
We assume that more similar wireless platforms exist, however, confirming this requires an extensive analysis of the respective firmware binaries.
Due to the popularity of \ac{ARM} and \emph{ThreadX}, we assume that there are further \fuzztool targets.

A platform that uses \ac{ARM} and \emph{ThreadX} and implements a wireless standard is 
\emph{Marvell Avastar} Wi-Fi~\cite{marvell2019}.
Moreover, the \emph{Huawei} baseband, as well as the \emph{Shannon} baseband in \emph{Samsung}
smartphones, are \ac{ARM}-based~\cite{amat}. 
\emph{Broadcom's} \mbox{Wi-Fi} chips are ARM-based, but the operating system is \emph{HNDRTE}~\cite{2017:googleprojectzero}.
We took a deeper look into the \emph{Raspberry Pi 3+/4} and \emph{Samsung Galaxy S9} Wi-Fi firmware and compared them to the Bluetooth firmware with known symbols.
We found that the \texttt{main} function in Wi-Fi and Bluetooth calls \texttt{\_tx\_initialize\_kernel\_enter}. Thus, both \emph{Broadcom} wireless stacks use \emph{ThreadX} for threading, timers, and events. Yet, Wi-Fi uses \emph{HNDRTE} functions instead of \emph{ThreadX} functions for memory management.

\subsection{Patching Bluetooth Vulnerabilities}
\label{ssec:patching}

\emph{Broadcom} Bluetooth chips are released with a fixed ROM image.
Patches are applied using a special Patchram mechanism~\cite{mantz2019internalblue}.
Each Patchram slot is temporarily stored in a remapped RAM section and consists of \SI{4}{\byte}. This is sufficient to insert a branch instruction to code stored in a regular RAM section.
The operating system applies device-specific patches during driver initialization.

\todo{maybe put flashpatch also here?}
Depending on the chip, there can be 128 or 256 Patchram slots.
This increasing number shows the need to be able to apply more patches.
Analysis of operating system patches reveals that 256 Patchram slots are by far not sufficient.  An overview is shown in \autoref{tab:patches}.
Moreover, the RAM area containing the code each patch jumps into is limited.
Overall, even recently released devices only allow for a few more patches.
Manufacturers like \emph{Apple}, who support devices for multiple years, cannot include all patches. For example, \cvea was fixed in \emph{iOS 12.4} on all devices except the \emph{iPhone 6}, which already uses all Patchram slots.

\emph{Broadcom} claimed \cveb would not be an issue despite producing a heap overflow.
Thus, we assume that \emph{Broadcom} only ships security updates for issues that are publicly known and that they consider exploitable.
The limited Patchram slots force them into this decision.
To this end, expanding the \fuzztool fuzz cases beyond zero-click attacks would likely result in further issues that \emph{Broadcom} would decide not to patch.

When initially finding \cvea, it was exploitable on any \emph{Broadcom} chip we tested.
Surprisingly, during responsible disclosure, \emph{Broadcom} stated that they knew about the issue since February 2018. We could confirm this because the \emph{Samsung Galaxy S10e} ROM contains a fix and has a compile date of April 2018. Interestingly, the most recent \emph{Cypress} evaluation board \emph{CYW20819} with firmware from May 2018 does not contain a fix.

Device manufacturers need to trust \emph{Broadcom} to include proper patches. One of the device manufacturers claimed that \emph{Broadcom} assured them the devices had been patched, despite being vulnerable in our tests. Dissecting and confirming patches at large scale is very hard for anyone besides \emph{Broadcom}. Binary diffing tools perform poorly on raw \ac{ARM} binaries, as correct function identification due to duplicate meanings in \emph{Thumb} mode at \SI{2}{\byte} offsets is challenging~\cite{polypyus}. Advanced graph analysis methods fail on this firmware because state-of-the-art disassemblers miss a significant amount of functions, thus, corrupting call graphs. Despite only differing in \ac{ARM} \emph{Cortex M3} versus \emph{M4}, having comparable compiler options, and similar hardware register locations, less than \SI{6}{\percent} of the functions could be identified in practice between the \emph{Nexus 5} firmware and the \emph{CYW20735} evaluation board firmware using \emph{BinDiff}~\cite{mantz2019internalblue}.

\begin{table}[!t]
\renewcommand{\arraystretch}{1.3}
\caption{Patchram slots used on various \emph{Broadcom} devices.}
\label{tab:patches}
\scriptsize
\centering
\begin{tabular}{|l|l|l|l|}
\hline
\textbf{Chip} & \textbf{Device} & \textbf{OS} & \textbf{Slots}\\
\hline
BCM4345B0 & iPhone 6 & iOS 12.4 & 128/128 \\
BCM4345C0 & Raspberry Pi 3+/4 & Raspbian Buster &128/128\\
BCM4345C1 & iPhone SE & iOS 12.4 & 127/128 \\
BCM4347B0 & Samsung Galaxy S8 & Android 9 &  254/256 \\
BCM4347B1 & iPhone 8/X/XR & iOS 13.4.1  &240/256 \\
BCM4375B1 & Samsung Galaxy S10/S10+ & Android 9 &  212/256\\
\hline
\end{tabular}
\vspace{-1em} 
\end{table}

\subsection{Memory Protection in Broadcom Chips}
\label{ssec:memprotect}

\emph{Broadcom} announced the introduction of  \emph{critical area access} memory protection to prevent attacks like \cvec. 
The idea is that special purpose registers, such as those for coexistence, can only be configured during device initialization and are locked afterward.
Despite reporting \cvec in August 2019, we did not see \emph{critical area access} as a patch in any firmware as of February 2020.
We assume that this feature is infeasible because the underlying \ac{ARM} chip is a \emph{Cortex M3} on chips prior to 2016 and
a \emph{Cortex M4} on newer chips~\cite{polypyus}, neither of which support such a feature.

After further questions to the \emph{Broadcom} security team about how and when \emph{critical area access} will be applied, we
finally saw something potentially related to this feature in \emph{iOS 13.4.1} and the March 2020
\emph{Samsung Android} release. Instead of protecting memory at the chip-level, the \ac{HCI}
commands to read and write memory are restricted, including the undocumented super duper peek poke command. After driver initialization, these
commands are blocked. While this helps against misusing \path{bluetoothd} to block the
Wi-Fi chip causing \ac{DoS}, it does not protect from over-the-air \ac{RCE} on the Bluetooth
chip and further escalation into the Wi-Fi chip.

\subsection{Heap Management in ThreadX}
\label{ssec:threadxheap}
\cvea, \cveb, and \cved exploit the heap structure in the underlying operating system. Patching this would secure 6.2 billion systems running \emph{ThreadX}. We proposed \emph{Express Logic} to integrate a heap sanitizer. 
As the \texttt{BLOC} structure contains fixed sizes, these checks run in constant time and could have fully mitigated our exploit technique and helped developers to detect vulnerabilities.
They responded that we are not the first to exploit the \emph{ThreadX} heap---a similar attack was published a few months before against \emph{Marvell Avastar} Wi-Fi chips~\cite{marvell2019}. Nonetheless, they do not plan to integrate any mitigation, stating that applications are responsible for secure heap access.

Despite this statement, the \emph{Samsung Galaxy S10e} performs a very basic heap check. 
We do not know whether \emph{Broadcom} or \emph{Express Logic} introduced it.
Crafting valid payloads is possible with the new check, but the payload needs to be adapted for each firmware version. This is already a requirement for all attacks that rely on calling functions and do not only write to special hardware registers.

\section{Related Work}
\label{sec:related}

In the following, we summarize existing work on wireless chip exploitation as well as Bluetooth fuzzing.

To the best of our knowledge, publicly available work on Bluetooth fuzzing only covers host implementations. Firmware has not been extensively fuzzed or systematically tested. Vendors might have non-public testing mechanisms.
Yet, the previously listed findings in wireless firmware show that vendors do not have sufficient techniques to prevent heap and buffer overflows.

So far, Bluetooth firmware research has been limited to extend chip functionality.
\emph{btlejack} builds on the documented \emph{Nordic Semiconductor} \ac{BLE} firmware~\cite{github:btlejack}. It supports passive and active \ac{MITM} attacks including \ac{BLE} 5 hopping.
In contrast, \emph{InternalBlue} is based on reverse-engineered \emph{Broadcom} chips~\cite{mantz2019internalblue}. While it does not support \ac{MITM} attacks, it can read and modify lower layer packets for both \ac{BLE} and classic Bluetooth. During the implementation of \emph{InternalBlue}, the authors manually detected a security issue on various \emph{Broadcom} chips. Despite the existing works on \emph{Nordic Semiconductor} and \emph{Broadcom} firmware,
there has not been any public, systematic security testing on these chips.

An over-the-air fuzzing on top of \ac{HCI} was implemented in~\cite{deepsec2011}. This black-box testing approach only detects crashes. These crashes might happen in the firmware, however, due to the implementation focusing on host layer protocols, crashes are most likely to happen in the operating system.
The remaining fuzzing implementations focus on the driver and operating system level and do not involve any over-the-air packets. For example, \texttt{syzkaller} supports fuzzing \ac{HCI} on Linux~\cite{syzkaller}. Moreover, \emph{kAFL} and its successors support fuzzing the Linux kernel~\cite{grimoire,kafl}.
Implementation faults in operating system components handling Bluetooth can lead to \ac{RCE} across various operating systems, as the \emph{Blueborne} attacks demonstrated~\cite{2017:blueborne}. Such---even wormable---escalations still exist in recent implementations as \emph{CVE-2020-0022} alias \emph{BlueFrag} shows~\cite{bluefrag}.

\emph{Broadcom's} Wi-Fi chips were initially exploited in 2017 by two independent research teams~\cite{2017:artenstein, 2017:googleprojectzero}. Heap exploitation was documented in~\cite{2017:googleprojectzero}, however, the heap is structured differently in the \emph{HNDRTE} operating system. Recently, new \emph{Broadcom} Wi-Fi vulnerabilities have been revealed~\cite{quarkslab2019}.

Other chipsets were also successfully exploited. The \emph{Marvell Avastar} Wi-Fi uses similar technologies as \emph{Broadcom} and had comparable heap vulnerabilities~\cite{marvell2019}. The author was using \path{afl-unicorn} for fuzzing~\cite{aflunicorn}, but did neither document the precise setup nor publish any source code.
The \emph{Intel} LTE stack is based on x86, and was successfully exploited despite memory protection mechanisms~\cite{guy2019}.
Moreover, the \emph{MediaTek} baseband exists in an \ac{ARM} and a MIPS variant and both were fuzzed based on the emulation of security-relevant protocol handlers~\cite{grassi,basesafe}.
\emph{Qualcomm} is using their own architecture and assembly for \ac{DSP}, \emph{Hexagon}, and implements various memory protection mechanisms as well as secure boot. Nonetheless, an over-the-air Wi-Fi buffer overflow exploit that escalates into the Linux kernel driver's memory allocation was found~\cite{defconqualcomm}.
Security analysis of the Wi-Fi firmware was done manually.

In general, emulation-based fuzzing is also supported by \emph{TriforceAFL}~\cite{triforceafl}. However, \emph{TriforceAFL} does not use \ac{QEMU} user-mode emulation like \fuzztool but full-system emulation.
Instead of adding hooks to the firmware, it modifies \ac{QEMU}.

In contrast to static program analysis and emulation-based fuzzing, \emph{LTEFuzz} performs over-the-air analysis on LTE and found vulnerabilities in various mobile devices and core network components~\cite{kim_ltefuzz_sp19}.
Moreover, \emph{SpikerXG} wirelessly fuzzes 2G on multiple smartphones in parallel, including a packet mutator using \emph{YateBTS}~\cite{hernandez2019basebads, yatebts}.
Such approaches are feasible for 2G and LTE, because open source projects like \emph{OpenAirInterface} and \emph{srsLTE} already implement a lot of common protocol features on \acp{SDR}~\cite{oai,srslte}.
For Bluetooth, there is currently no comparable implementation. Moreover, over-the-air analysis is often unable to determine the precise causes of crashes.


\section{Conclusion}
\label{sec:conclusion}

In this paper, we demonstrate several security problems originating from Bluetooth \ac{RCE}---ranging from issues with the Bluetooth specification to broken driver implementations in various operating systems. Our findings unveil the possibility to escalate beyond the Bluetooth circuit boundaries: attackers may take control of the chip over-the-air and, from there, start disturbing Wi-Fi and LTE communications or even crash the entire smartphone.

We create \fuzztool, a tool for non-wireless fuzzing of wireless firmware in an emulated environment. \fuzztool restarts emulation from snapshots of the device's physical state after frame reception. As it brings fuzzing speed to an unprecedented level, it can be attached to complex operating systems and find full-stack bugs.

Emulation allows understanding \ac{RCE} vulnerabilities that do not immediately cause a crash but are potentially dangerous.
The findings covered in this paper got us in contact with further chip manufacturers, confirming that there is high interest in and awareness of technologies that allow testing wireless implementations and help fixing vulnerabilities.

The vulnerability patching issues of \emph{Broadcom} Bluetooth chips highlight the importance of building sustainable and secure update mechanisms. 
We found the overall responsible disclosure process quite alarming. One of our attacks, \cvea, was internally discovered by \emph{Broadcom} in February 2018, but when we informed them about our findings in April 2019, our \ac{PoC} was working on all \emph{Broadcom} chips we had access to. Usually, until a Bluetooth chip becomes available on off-the-shelf devices, it is at least one year old.
Due to the patching mechanism constraints and ease of analyzing patches, \emph{Broadcom} cannot patch all vulnerabilities on older chips. Each patch comes with a high risk of leaking a vulnerability.
Despite monthly contact with \emph{Samsung}, a fix for \cvea took until mid September 2019 on the \emph{Samsung Galaxy S8}, which is comparably well-supported.

Despite failing to fix Bluetooth firmware vulnerabilities, mobile operating systems integrate Bluetooth into critical components.
With the overall presence of Bluetooth, even worms spreading wirelessly become feasible.
Recent mobile operating systems do not reset and disable Bluetooth properly, even though they suggest to users that they do. The advice to turn off Bluetooth when not needed is insufficient. Always being connected is a very alarming trend regarding over-the-air attacks. 
Ideally, this trend can be reversed in the future, thus, giving back control over wireless technologies to the users.

\section*{Acknowledgments}

We thank \emph{Apple}, \emph{Broadcom}, \emph{Cypress}, \emph{Express Logic}, \emph{Fitbit}, \emph{Google}, and \emph{Samsung} for handling the responsible disclosure requests,
and Ren\'{e} Mayrhofer for assisting us in the responsible disclosure process.
Moreover, we thank Dennis Heinze for porting \emph{InternalBlue} to \emph{iOS} and testing \cvec on various \emph{iPhones}, Dennis Mantz for testing it on the \emph{iPhone X}, and Michael Sp\"ork for the \ac{BLE} expertise.
We also thank Lars Almon, Oliver P\"ollny, Bianca Mix, Tim Walter, Dominik Maier, and Teal Starsong for proofreading and Nils Ole Tippenhauer for shepherding this paper.

This work has been funded by the German Federal Ministry of Education and Research and the Hessen State Ministry for Higher Education, Research and the Arts within their joint support of the National Research Center for Applied Cybersecurity ATHENE.

\section*{Availability}

\fuzztool is publicly available on \url{https://github.com/seemoo-lab/frankenstein}.


\bibliographystyle{plain}
\bibliography{bibliographies}


\end{document}
